\documentclass[journal,draftcls,twocolumn,10pt,twoside]{IEEEtran}
\normalsize

\ifCLASSINFOpdf
\else
\fi

\usepackage[cmex10]{amsmath}
\usepackage[caption=false,font=footnotesize]{subfig}

\usepackage{tikz}
\usepackage{pgfplots}
\usepackage{multirow}
\begin{document}
%
\title{On the Probability of Partial Decoding in Sparse Network Coding}
%
%
%

\author{Hadi~Sehat\IEEEmembership{}
        and~Peyman~Pahlavani\IEEEmembership{}}%

%
%

\markboth{IEEE Transactions on Networking}%
{paper}
%



\maketitle

\begin{abstract}
Sparse Network Coding (SNC) has been a promising network coding scheme as an improvement for Random Linear Network Coding (RLNC) in terms of the computational complexity. However, in this literature, there has been no analytical expressions for the probability of decoding a fraction of source messages after transmission of some coded packets. In this work, we looked into the problem of the probability of decoding a fraction of source messages, i.e., partial decoding, in the decoder for a system which uses SNC. We exploited the Principle of Inclusion and Exclusion to derive expressions of partial decoding probability. The presented model predicts the probability of partial decoding with an average deviation of 6~\%. Our results show that SNC has a great potential for recovering a fraction of the source message, especially in higher sparsity and lower Galois Field size. Moreover, to achieve a better probability of partial decoding throughout transmissions, we define a sparsity tuning scheme that significantly increases the probability of partial decoding. Our results show that this tuning scheme achieves a 16~\% improvement in terms of probability of decoding a fraction of source packets with respect to traditional SNC.
\end{abstract}


\ifCLASSOPTIONpeerreview
 \begin{center} \bfseries EDICS Category: 3-BBND \end{center}
\fi
%
\IEEEpeerreviewmaketitle

\section{Introduction}
%
%
%
%
\IEEEPARstart{R}{andom} Linear Network Coding (RLNC), is a network coding scheme, in which coded packets are constructed by random linear combinations of source packets~\cite{RLNC}. Sparse Network Coding (SNC) is shown to be a major improvement for RLNC, reducing its high computational complexity with the expense of larger communication overhead~\cite{Garrido}. In both RLNC and SNC schemes, the encoder sends $n$ source packets, i.e., generation, to a destination, i.e., the decoder. The decoder decodes the whole message after exactly $n$ linearly independent coded packets are received~\cite{RLNC}~\cite{Tassi2018}. This decoding procedure leads to long delays as huge number of transmissions are required to decode the whole generation. In order to overcome this property, rank-deficient decoding was proposed in~\cite{RDD} to recover a subset of source packets from fewer than $n$ received coded packets. In this literature, probability of partial decoding is the probability of decoding at least $x \leq n$ source packets when $m$ coded packets are received, where $m$ can be smaller, equal to or greater than $n$~\cite{RLNCADD}.

The problem of decoding a fraction of source packets in a network coding scheme has been studied as the concept of secure network coding, e.g.,~\cite{RLNCSEC}. In this scenario, the security criteria is that a potential eavesdropper does not have any information about the source packets. This security criteria shows that if we decrease the probability of partial decoding, the system is more secure~\cite{Tassi2018}. Other than security concerns, the probability of partial decoding is of importance in real-time systems and live streams. In traditional RLNC, the decoder must receive exactly $n$ linearly independent coded packets to recover the whole generation which leads to huge delay if $n$ is large or the channel error is high. However, by using rank-deficient decoding, the decoder can recover a fraction of source packets while less than $n$ coded packets are received, hence reducing the delay per packet throughout the transmission~\cite{RLNCADD}. 

As far as our knowledge, there is only one paper on the probability of partial decoding of SNC. A recent work in this literature is~\cite{Amir}, which provides a Markov Chain model to derive the aforementioned probability. However, this work is limited for high field size assumption where the authors used Markov Chain model. If the system variables change, a new set of Markov Chain analysis is needed to derive a performance model. 

Tunable Sparse Network Coding (TSNC) was presented in ~\cite{Feizi}, ~\cite{GarridoTuning} to achieve better performance in terms of completion time and complexity. Completion time is the number of transmissions until the last packet is decoded in a generation. In this work, the author did not provide any model for partial decoding. A proper tunable approach to increase the probability of partial decoding is also missed in the literature.

Our contribution in this paper is as follows.
\begin{enumerate}
	\item Unlike~\cite{Amir}, we provide an approximation for the probability of partial decoding for an SNC scheme in general form. Our model predicts the probability of decoding at least $x$ source packets after a specific number of transmissions. An approximation for the probability of decoding exactly $x$ source packets when $m$ coded packets are transmitted is also derived.
	\item We derived approximate expressions for the probability of partial decoding and for a transmission channel with an erasure probability.
	\item We propose a tuning scheme for SNC network to enhance the performance of SNC scheme in terms of partial decoding probability. We prove that this tuning scheme has the minimum Average Decoding Delay (ADD)~\cite{Amir} for a system using SNC scheme.
	\item We validated our model for the probability of partial decoding using simulation results. The results show a maximum Mean-Absolute-Percentage-Deviation (MAPD)~\cite{MAPD} of 6~\% between simulation and theoretical results for different values of the number of source packets, Galois Field size and sparsity.
	\item We tested our tuning scheme using simulation techniques. The results show an average of 16~\% reduction of ADD for our tuning scheme with respect to the traditional SNC scheme throughout the transmission.
\end{enumerate} 

The rest of this paper is organized as follows. Section~\ref{related} includes the related work. In section~\ref{problem}, we formulate our problem and introduce our system model. In section~\ref{prob}, we derive the probability distribution for decoding a fraction of source packets. In section~\ref{tuning}, we present our tuning scheme. Section~\ref{res} presents our results and finally section~\ref{conclusion} includes the conclusion of this work and future research directions.

\section{Related Work}
\label{related}
The probability of decoding a fraction of source packets in RLNC scheme has been a major research topic, in context of both performance~\cite{RLNCADD} and security~\cite{RLNCSEC}.  The authors of~\cite{RLNCSEC} showed an upper bound for the possibility of decoding a fraction of source packets, while in~\cite{RLNCADD}, the authors derived exact expressions for the probability of partial decoding for RLNC. Unfortunately, none of these works can be extended to SNC scheme. In~\cite{Amjad}, these expressions were used to study the security of RLNC in a multi-relay network. The authors of~\cite{BinaryRLNC} also found an exact expression for the probability of partial decoding in systematic RLNC. However, their analysis is only valid for Binary Galois Field and also can not be extended to SNC.

Rateless codes such as LT and Raptor codes, can be considered as a binary implementation of SNC~\cite{LT}~\cite{Raptor} and partial decoding probability has been a major research topic in the literature of rateless codes. To mention a few, the authors of~\cite{Suj} designed an algorithm for an optimal recovery rate, i.e. partial decoding probability in LT codes. However, the results of this work are asymptotically optimal and may only be employed for infinite number of source packets. In~\cite{LTanal}, the authors provided a probability analysis for decoding a fraction of source packets based on the structure of the received coded packets. However, their analysis can not provide any probability for partial decoding, where the exact structure of the received coded packets are unknown. The authors of~\cite{Talari} and~\cite{kim}, proposed algorithms to increase the probability of partial decoding for rateless codes. However, these works only increase the probability of partial decoding in the specific stages of the whole transmission and also can not be extended to non-binary Galois Fields. Also these algorithms are based on current coded packets received by the decoder and require huge computation overhead while transmitting coded packets.

Another interesting research is Instantly Decodable Network Coding~(IDNC)~\cite{IDNC}. In this scheme, the packets are sent in a way that an instant decoding of at least one of the source packets is guaranteed. However, the analysis on the partial decoding probability of this family of codes, such as~\cite{neda} and~\cite{arxiv}, is only valid for binary Galois Field and the presence of feedback in the system. Moreover, the IDNC family of codes heavily rely on feedbacks, though in our work, the system is considered to be feedback-free.

The authors of~\cite{perp} and~\cite{pey} introduced and analyzed an improvement for the SNC scheme called perpetual coding. However, this coding scheme is not completely random and uses a structured type of coding to send packets, and the analysis on this coding scheme can not be extended to random SNC scheme, which is the scope of this paper.

\section{Problem Statement}
\label{problem}
We consider a sender, i.e., encoder, which wants to send $n$ source packets $\{p_{1},p_{2},...,p_{n}\}$ to a destination node, i.e., decoder in an error-free channel. The decoder receives $m$ coded packets, where each coded packet is generated by $C=\sum\limits_{i=1}^{n}c_{i}p_{i}$ and the coefficients $\{c_{i},...,c_{n}\}$ are generated from a Galois Field $GF(q)$ using the following probability distribution:
\begin{equation}
P(c_{i} = v) = \begin{cases}
1-p\qquad if\; v\; =\; 0\; ; \\
\frac{p}{q-1}\quad Otherwise 
\end{cases}
\label{eq:p}
\end{equation}
where $p \leq 1-\frac{1}{q}$, RLNC is achieved when $p = 1-\frac{1}{q}$. In this notation, $1-p$ denotes the sparsity of the system~\cite{p}.
 
The decoder stores the coefficients of the received coded packets into matrix $M$, called decoding matrix. The decoding matrix $M$ is a member of  $F_{q}^{m,n}$, the set of all $m$-by-$n$ matrices that can be generated over $GF(q)$. The decoder decodes all source packets if and only if the decoding matrix contains exactly $n$ linearly independent rows, i.e., the rank of the decoding matrix is $n$. 

Although the network coding strategies including RLNC are used in different network topologies, there are still no recoding mechanism for SNC in multi-relay networks. There have been some research on the recoding aspect of SNC such as~\cite{recode}, however, the proposed methods are not guaranteed to preserve the sparsity of the packets throughout the transmission. Hence, our analysis in this work is limited to a single-hop network.

In this work, we approximated the SNC scheme in a way that in each coded packet, there are exactly $n\cdot p$ non-zero coefficients in random positions, i.e., the sparsity of each row is fixed. The same approach has been followed by~\cite{Amir} and~\cite{Feizi}. This approximation is valid in higher number of source packets and the results show that the derived model is also valid for general SNC scheme. We denote this model as constrained SNC in this paper.

In this paper, we suppose $i^{th}$ source packet as $e_{i}$, which is a vector of length $n$. The $i^{th}$ element of this vector is 1 while all other elements are zero. By this notation, the source packet $i$ is decoded if and only if $e_{i}\in Row(M)$, where $Row(M)$ is the row space of $M$~\cite{RLNCADD}.

Although the derived equations and the proposed tuning scheme are based on the error-free channel, we have also derived he probability of partial decoding for an erasure channel in section~\ref{prob}.

\section{Probability Analysis}
\label{prob}

In this paper, we are interested in derivation of the probability distribution for decoding at least $x \leq n$ source packets, given $m$ coded packets are received by the decoder, where $m$ can be smaller than, equal to or bigger than $n$. This probability is denoted by $P(X\geq x|M=m)$. In order to find this probability, we decompose our analysis into two parts.
\begin{enumerate}
	\item The probability that at least $x$ source packets are decoded, given $r$ out of $m$ received coded packets are linearly independent. This probability is denoted by $P(X\geq x|R=r,M=m)$.
	\item The probability that $r$ out of $m$ received coded packets are linearly independent. This probability is denoted by $P^{r}_{m}$.
\end{enumerate}

It is easy to see that these probabilities relate to each other as~\cite{RLNCADD}: 
\begin{equation}
P(X\geq x|M=m) = \sum_{r=x}^{m}P(X\geq x|R=r,M=m)P^{r}_{m}.
\label{eq:start}
\end{equation}

The probability that $r$ out of $m$ coded packets are linearly independent has been found for SNC scheme in~\cite{my}, where $P^{r}_{m}$ is derived by using a recursive equation. In this section, we modify the equations given in aforementioned paper to derive an equation for $P^{r}_{m}$. According to~\cite{my}, the probability distribution for rank of a sparse matrix $m$-by-$n$ matrix, where $m \leq n$ and the entries of the matrix are distributed by~\eqref{eq:p} in $GF(q)$, is as follows:
\begin{equation}
\begin{split}
&P_{m}^{m} = P_{m-1}^{m-1}\frac{P_{m}^{full}}{P_{m-1}^{full}}\; ; \\
&P_{m}^{r} = P_{m-1}^{r}\Big(1 - \frac{P_{r+1}^{full}}{P_{r}^{full}}\Big) + P_{m-1}^{r-1}\frac{P_{r}^{full}}{P_{r-1}^{full}}\; ;  \\
&P_{m}^{0} = P_{m-1}^{0} \Big(1-\frac{P_{1}^{full}}{P_{0}^{full}}\Big)\; ; \\
&\forall\; r : 0 < r < m\quad and\quad
P_{0}^{full} = P_{0}^{0} = 1 \\
\end{split}
\label{eq:2}
\end{equation}

where $P^{full}_{i}$ is the probability that a matrix with $i$ rows, where its elements are distributed as~\eqref{eq:p}, is full rank~\cite{RankLT}. This probability is equal to~\cite{my}:
	\begin{equation}
	P^{full}_{i} = \Pi_{k=1}^{i}(1-p_{k})^{n_k}
	\label{eq:full}
	\end{equation}

where $p_{k}$ and $n_{k}$ are defined as follows:

\begin{equation}
\begin{split}
&p_{k} = \big(\frac{q-1}{q}(1-\frac{qp}{q-1})^{k} + \frac{1}{q}\big)^{n}\; ; \\
&n_{k} = {i\choose k}(q-1)^{k}.
\end{split}
\end{equation}

We write~\eqref{eq:2} in matrix form as in~\eqref{eq:final1}. 

\begin{figure*}[!t]
	\normalsize
	\setcounter{equation}{5}
	\begin{equation}
\begin{bmatrix}
P^{0}_{m} \\
P^{1}_{m} \\
\vdots \\
P^{m}_{m}
\end{bmatrix} = 
\begin{bmatrix}
1-P^{full}_{0} &  &  &  &  \\
P^{full}_{0}       & 1-P^{full}_{1}  & & & \\
&\ddots & \ddots &  &  \\
& & P^{full}_{m-1} & 1 - P^{full}_{m} &
\end{bmatrix}^{m-1}
\begin{bmatrix}
1 \\
0 \\
\vdots \\
0
\end{bmatrix}
= A^{m-1}e_{1}.
\label{eq:final1}
	\end{equation}
	\setcounter{equation}{6}
	\hrulefill
	\vspace*{4pt}
\end{figure*}

\subsection{Probability of partial decoding in error-free channel}

This subsection is devoted to the derivation of the probability of decoding at least $x$ source packets, given the number of rows of the decoding matrix and the rank of the decoding matrix, i.e., $P(X \geq x|M=m, R=r)$. Before we start, we need to define the mathematical tools used in this section.

\textbf{Principle of Inclusion and Exclusion.} Given a set $A$, let $f$ be a real valued function defined for all sets $S$, $J\subseteq A$. If $g(S) = \sum_{J: J\supseteq S}f(J)$, then $f(S) = \sum_{J: J\supseteq S}(-1)^{|J/S|} g(J)$~\cite{prince}.

\textbf{The general form of Binomial Coefficients.} The general form of binomial coefficients for real numbers is as follows~\cite{fowler}:
\begin{equation}
{n\choose k} = \frac{\Gamma(n+1)}{\Gamma(k+1)\Gamma(n-k+1)}.
\label{eq:gen}
\end{equation} 

In order to find the probability $P(X\geq x|R=r,M=m)$, first, we find the probability of recovering exactly $x$ source packets, denoted by $P(X = x|R=r,M=m)$. In order to find this probability, we first propose two lemmas.

\textbf{Lemma 1.} The number of possible decoding matrices when the decoder receives $m$ coded packets generated from $n$ source packets in a constrained SNC transmission is equal to:
\begin{equation}
N = \big(\frac{\Gamma(n+1)}{\Gamma(np+1)\Gamma(n-np+1)}(q-1)^{np}\big)^{m}.
\label{eq:lem1}
\end{equation}

\textit{Proof.} In a constrained SNC scheme where the probability that each coefficient is non-zero is equal to $p$, each coded packet has an average of $np$ non-zero coefficients. Therefore, on average, each coded packet has ${n\choose np}(q-1)^{np}$ possible instances and $\big({n\choose np}(q-1)^{np}\big)^{m}$ possible decoding matrices. Since $np$ may have a non-integer value, we substitute ${n\choose np}$ with its equivalent using~\eqref{eq:gen} to derive~\eqref{eq:lem1}.

\textbf{Lemma 2.} The number of decoding matrices when the decoder receives $m$ coded packets generated from $n$ source packets in a constrained SNC transmission, and $x$ source packets are decoded is equal to

\begin{equation}
\begin{split}
&N_{x} = \\
& \begin{cases}
\big(\frac{\Gamma(n+1)}{\Gamma(np_{x}+1)\Gamma(n-np_{x}+1)}(q-1)^{np_{x}}\big)^{m-x} \qquad\ if\; x \leq m\; ; \\
0 \qquad\qquad\qquad\qquad\qquad\qquad\qquad\qquad\, Otherwise
\end{cases} \\
\end{split}
\end{equation}

where

\begin{equation}
P_{x} = \sum_{l=0}^{x} \frac{np - l}{n}{x\choose l}\frac{\Gamma(np-l+1)\Gamma(n+1)}{\Gamma(n-l+1)\Gamma(np+1)}.
\label{eq:pd}
\end{equation}

\textit{Proof.} Suppose $x$ source packets are decoded, hence, there are $m-x$ coded packets in the decoding matrix. Since the decoder uses Gauss-Jordan elimination to decode the source packets, the coefficients of the decoded source packets is changed to zero in the remaining $m-x$ coded packets. Therefore, the sparsity of the remaining coded packets is not equal to $1-p$. Let's denote the sparsity of the remaining coded packets as $1-P_{x}$. In order to calculate $P_{x}$, we use the average of sparsity of all the coded packets.

The probability that $l$ out of $x$ decoded source packets are contained in one of the remaining coded packets is equal to:
\begin{equation}
P_{l,x} = \frac{{n-l\choose np-l}}{{n\choose np}}{x\choose l}.
\label{eq:pl}
\end{equation}

The probability that the sparsity of one of the remaining coded packets is $1-\frac{np-l}{n}$ is equal to~\eqref{eq:pl}. Hence, the expected probability of having non-zero coefficient in the remaining $m-x$ coded packets is equal to:
\begin{equation}
P_{x} = \sum_{l=0}^{x}\frac{np-l}{n}P_{l,x}.
\end{equation}   

By substituting the general form of ${n\choose k}$ using~\eqref{eq:gen}, we get~\eqref{eq:pd} for the probability of having non-zero elements in the remaining coded packets. Therefore, the total number of decoding matrices for which $x$ source packets are decoded, is found using~\eqref{eq:lem1}, by substituting $m$ with $m-x$ and $p$ with $P_{x}$. 

\textbf{Theorem 1.} Given a decoder have received $m$ sparse coded packets with sparsity $1-p$, generated from from $n$ source packets, the probability that exactly $x$ source packets are decoded is given by

\begin{equation}
\begin{split}
&P(X = x|R=r,M=m) = {n\choose x} \sum_{j=0}^{m-x}\Big[{n-x\choose j} \\
&(-1)^{j}\frac{\big(C_{x,j}\cdot (q-1)^{np_{x+j}}\big)^{m-x-j}P_{m-x-j}^{r-x-j}}{nP_{m}^{r}}\Big].
\end{split}
\label{eq:final2}
\end{equation}

where
\begin{equation}
C_{x,j} = \frac{\Gamma(n+1)}{\Gamma(np^{x +j}+1)\Gamma(n-np_{x+j}+1)}.
\label{eq:C}
\end{equation}

\textit{Proof.} In appendix 1.

\textbf{Corollary 1.} For a system using constrained SNC scheme to generate coded packets, the probability of decoding at least $x$ source packets given the decoding matrix has $m$ rows and rank $r$,  is equal to:

\begin{equation}
\begin{split}
&P(X \geq x|R=r,M=m) = \sum_{i=x}^{r} {n\choose i} \sum_{j=0}^{m-i}\Big[{n-i\choose j} \\
&(-1)^{j}\frac{(C_{i,j}\cdot (q-1)^{np_{p}^{i+j}})^{m-i-j}P_{m-x-j}^{r-x-j}}{nP_{m}^{r}}\Big].
\end{split}
\label{eq:final3}
\end{equation}

\textit{Proof.} By definition, $P(X \geq x|R=r,M=m) = \sum\limits_{i=x}^{r} P(X = i|R=r,M=m)$, substituting in~\eqref{eq:final2} gives the result.

\textbf{Corollary 2.} For a system using constrained SNC scheme to generate coded packets, the probability of decoding at least $x$ source packets given the decoding matrix has $m$ rows is equal to:
\begin{equation}
\begin{split}
& P(X\geq x|M=m) = \\
&\sum_{r=0}^{m}\Big[(A^{m-1}e_{1})_{r}
\sum_{i=x}^{r} {n\choose i} \sum_{j=0}^{m-i}{n-i\choose j}(-1)^{j} \\
& \frac{\big(C_{i,j}. (q-1)^{np_{i+j}}\big)^{m-i-j}P_{m-x-|J^{\prime}|}^{r-x-|J^{\prime}|}P_{m-x-j}^{r-x-j}}{nP_{m}^{r}}\Big] 
\end{split}
\label{thefinal}
\end{equation}

where $(A^{m-1}e_{1})_{r}$ denotes $r^{th}$ row of the result matrix $A^{m-1}e_{1}$ and $C_{x,j}$ and $N$ are given by~\eqref{eq:C} and~\eqref{eq:lem1} respectively. 

\textit{Proof.} Using~\eqref{eq:start} and substituting in~\eqref{eq:final3}~\eqref{eq:final1} gives the result.

\textbf{corollary 3.} For a system using constrained SNC scheme to generate coded packets, the probability of decoding exactly $x$ source packets given the decoding matrix has $m$ rows is equal to:
\begin{equation}
\begin{split}
&P(X=x|M=m) = \sum_{r=0}^{m}\Big[(A^{m-1}e_{1})_{r}{n\choose x} \\ &\sum_{j=0}^{m-x}{n-x\choose j} (-1)^{j}\frac{\big(C_{x,j}. (q-1)^{np_{x+j}}\big)^{m-x-j}P_{m-x-j}^{r-x-j}}{NP_{m}^{r}}.
\end{split}
\label{eq:c3}
\end{equation}
\textit{Proof.} Using~\eqref{eq:start} and substituting in~\eqref{eq:final2} and~\eqref{eq:final1} gives the result.

\subsection{Probability of partial decoding in a erasure channel}

Consider the channel between the sender and the receiver to have an erasure probability equal to $\varepsilon$, i.e., a coded packet is received by the receiver with probability $1-\varepsilon$. In this case, the total number of received packets by the receiver after $m$ transmissions is equal to $m(1-\varepsilon)$. Therefore, the probability of partially decoding $x$ source packets is equal to:
\begin{equation}
\begin{split}
&P(X=x|M=m) =\sum_{r=0}^{m(1-\varepsilon)} \Big[(A^{m(1-\varepsilon)-1}e_{1})_{r} \\
& {n\choose x}\sum_{j=0}^{m(1-\varepsilon)-x}{n-x\choose j} (-1)^{j} \\
& \frac{\big(C_{x,j}. (q-1)^{np_{x+j}}\big)^{m(1-\varepsilon)-x-j}P_{m(1-\varepsilon)-x-j}^{r-x-j}}{NP_{m(1-\varepsilon)}^{r}}\Big]. 
\end{split}
\label{eq:error}
\end{equation}

\section{Tuning Scheme}
\label{tuning}

In order to present the tuning scheme for SNC, we use the ADD as the performance measure. ADD is defined as the expected number of transmissions required to decode a source packet form a generation~\cite{Amir}. We choose ADD as the performance measure because lower ADD indicates lower number of transmissions are required to recover each source packet. As the scope of this paper is to propose methods to increase the chance of decoding source packets by less transmission of coded packets, we can focus on a scheme to lower the ADD of the transmission.

 In this section, we first formulate ADD in terms of the sparsity of the system and then use this formulation in order to tune the sparsity of the system to achieve the minimum ADD. Using Corollary~3 in Section~\ref{prob}, we calculate the average number of decoded source packets after $m$ transmissions as shown by the following theorem.

\textbf{Theorem 2.} The average number of decoded source packets after $m$ successful transmissions in a system using constrained SNC scheme based on~\eqref{eq:p} is as follows.
\begin{equation}
\begin{split}
&E(X|M=m) = \sum_{r=0}^{m}\Big[(A^{m-1}e_{1})_{r}\sum_{x=0}^{n}x{n\choose x} \\
& \sum_{j=0}^{m-x}{n-x\choose j}(-1)^{j}\frac{\big(C_{x,j}. (q-1)^{np_{x+j}}\big)^{m-x-j}P_{m-x-j}^{r-x-j}}{np_{m}^{r}}\big].
\end{split}
\label{eq:theo2}
\end{equation}

\textit{Proof.} The probability of decoding exactly $x$ source packets after $m$ transmissions is given by~\eqref{eq:c3}. Computing the average of this equation over $x$ gives the result.

In order to propose our tuning scheme, we first propose two lemmas.

\textbf{Lemma 4.} Average ADD of a system using constrained SNC scheme is equal to:

\begin{equation}
E(ADD) = \sum_{m=0}^{m_{0}} m.\big(E(X|M=m) - E(X|M=m-1)\big)
\label{eq:ADD}
\end{equation}

where $m_{0}$ is the minimum value of $m$ for which $E(X|M=m_{0}) \simeq n$, i.e., the number of transmissions to decode the whole generation.

\textit{Proof.} If packet $i$ is expected to be decoded after $m_{i}$ transmissions, ADD can be formulated as~\cite{Amir}:

\begin{equation*}
ADD = \frac{\sum\limits_{i} m_{i}}{n}.
\end{equation*}

In other words, if the expected number of source packets that are decoded exactly after $m_{i}^{th}$ successful transmissions is denoted as $d_{i}$, the ADD can be formulated as: 

\begin{equation*}
E(ADD) = \frac{\sum\limits_{i=0}^{k} m_{i}d_{i}}{n}
\end{equation*}

where $k$ is the total number of transmissions until all of the source packets are decoded, in order to calculate $d_{i}$, we use~\eqref{eq:theo2}. It is easy to see that $d_{i} = E(X|M=m_{i}) - E(X|M=m_{i}-1)$, completing the proof.

\textbf{Lemma 5.} In order to achieve the minimum ADD in our system, we choose the sparsity of $i^{th}$, i.e., $p_{m}$, in a way that $E(X|M=i)$ is maximum for the given $p_{i}$. 

\textit{Proof.} We can rewrite~\eqref{eq:ADD} as follows.

\begin{equation}
\begin{split}
&E(ADD) \\
&= E(X|M=1) - E(X|M=0) + ... \\
& + m_{0}.E(X|M=m_{0}) - m_{0}.E(X|M=m_{0}-1)\\
&=  m_{0}.n-E(X|M=0) - E(X|M=1) -  \\
&E(X|M=2) -... -E(X|M=m_{0}-1)\\ 
&= m_{0}.n - \sum_{i=0}^{m_{0}-1} E(X|M=i). \\
\end{split}
\label{eq:ADD2}
\end{equation}

The value of $E(X|M=m)$ is bounded by $n$, so there exists a $p_{m_{0}}$ and $m_{0}$ for which the value of $E(X|M=m_{0}) = n$ where there is no value for $p_{m^{\prime}}$ and $m^{\prime} < m_{0}$ for which $E(X|M=m^{\prime}) = n$. In this notation, $m_{0}$ denotes the minimum value of $m$ for which the whole generation can be decoded. Our algorithm chooses a $p_{i}$ at each step $i$ for which $E(X|M=i)$ is maximum, hence, this algorithm always chooses the mentioned $p_{m_{0}}$ and $m_{0}$.
Therefore, our algorithm ensures that the value of $m_{0}$ is minimum and all values of $E(X|M-n)$ are maximum. Therefore, according to~\eqref{eq:ADD2}, the expected value of ADD for the system which uses the algorithm is its minimum.

In the proof of lemma~5 we also proved that we choose $p_{m}$ in a way that $E(X|M=i)$ is maximum for the given $p_{i}$, the total number of required transmissions to decode the whole generation is minimum.

By using~\eqref{eq:theo2}, we find a value $p_{i}$ for $i^{th}$ coded packet, where $E(X|M=m_{i})$ is maximum for sparsity $1-p_{i}$. Then, we use following lemma to calculate the sparsity of $i^{th}$ coded packet.

\textbf{Lemma 6.} To achieve the minimum ADD, we set the value $p$ in~\eqref{eq:p} for $i^{th}$ coded packet as follows:

\begin{equation}
p = i.p_{i} - (i-1)p_{i-1}.
\label{Pprime}
\end{equation}

\textit{Proof.} Before sending the $i^{th}$ coded packet, the sparsity of decoding matrix is equal to $1-p_{i-1}$. The $i^{th}$ coded packet must be sent in a way that after sending this packet, the sparsity of the decoding matrix become $1-p_{i}$. The sparsity of the decoding matrix is equal to average sparsity of its rows, hence, we have:

\begin{equation*}
\frac{p+ (i-1)p_{i-1}}{i} = p_{i}
\end{equation*} 

which leads to~\eqref{Pprime}.

The proposed tuning scheme calculates the value of $p$ for each coded packet using~\eqref{Pprime} prior to the transmission. This algorithm would prevent any additional computational complexity during the transmission. Hence the minimum value of ADD and the number of transmissions can be achieved by a pre-processing in the source node.

\section{Results}
\label{res}

In order to validate our results, we used KODO library~\cite{KODO} to simulate fully-random SNC scheme based on equation~\ref{eq:p} in a one-hop system, consisting of an encoder and a decoder. In this simulation campaign, we carried out 50000 independent experiments and derived the number of decoded packets after $m$ coded packets are received in the decoder. The expected number of decoded source packets is considered as the average value of the aforementioned measure in all experiments.

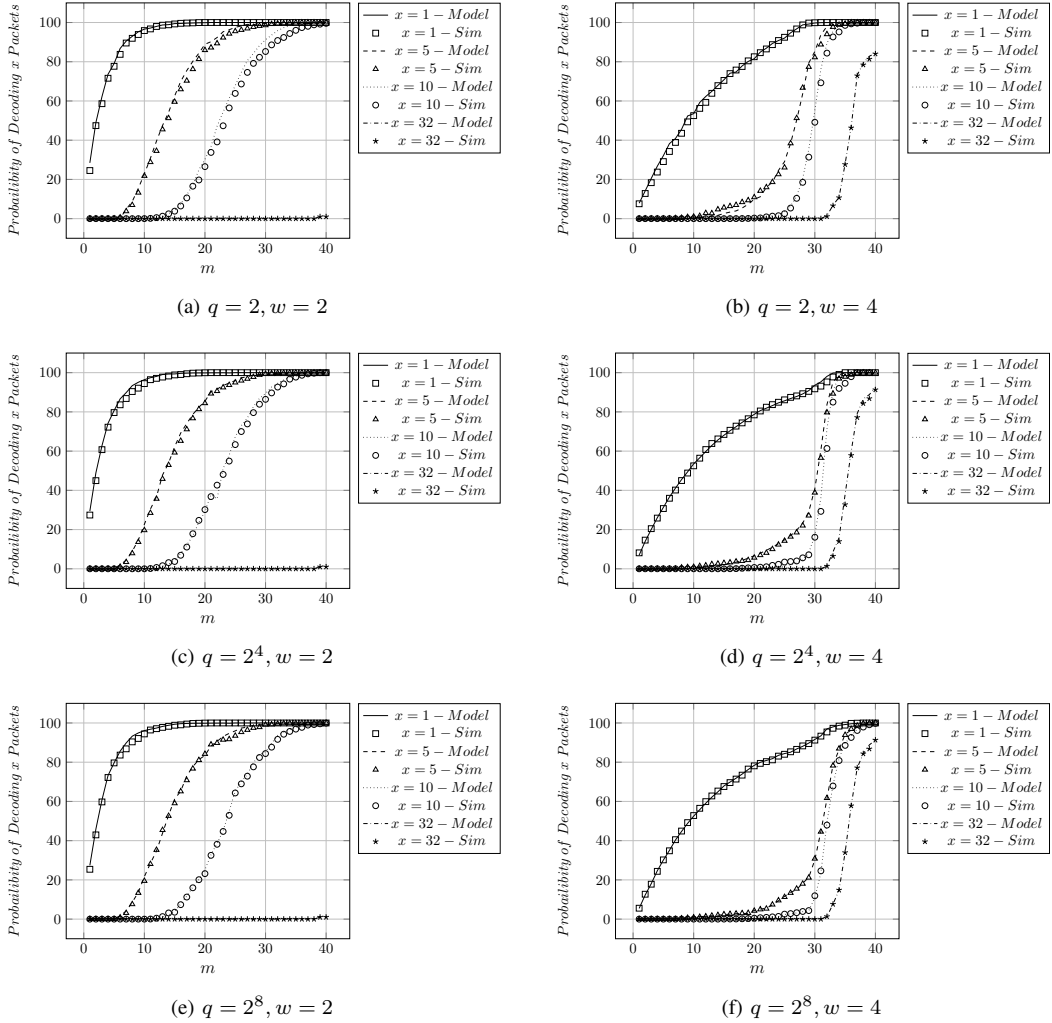
\begin{figure*}[!t]
	\centering
	\subfloat[$q=2, w=2$]{\begin{tikzpicture}
		[scale= 0.55]
		\begin{axis}[
		xlabel= $m$,
		ylabel=$Probailibity\; of\; Decoding\; x\; Packets$,
		xlabel style={font=\large},
		grid=major,
		legend pos= outer north east]
		
		\addplot[  black] plot coordinates {
			(1, 28.45)
			(2, 49.04)
			(3, 62.2)
			(4, 73.6)
			(5, 79.06)
			(6, 86.76)
			(7, 90.54)
			(8, 92.94)
			(9, 95.28)
			(10, 96.63)
			(11, 97.17)
			(12, 98.02)
			(13, 98.75)
			(14, 98.92)
			(15, 99.45)
			(16, 99.52)
			(17, 99.67)
			(18, 99.75)
			(19, 99.87)
			(20, 99.9)
			(21, 100)
			(22, 100)
			(23, 100)
			(24, 100)
			(25, 100)
			(26, 100)
			(27, 100)
			(28, 100)
			(29, 100)
			(30, 100)
			(31, 100)
			(32, 100)
			(33, 100)
			(34, 100)
			(35, 100)
			(36, 100)
			(37, 100)
			(38, 100)
			(39, 100)
			(40, 100)		
		};
		\addlegendentry{$x=1 - Model$}
		\addplot[only marks,mark=square,black] plot coordinates {
			(1, 24.53)
(2, 47.43)
(3, 58.62)
(4, 71.58)
(5, 77.68)
(6, 83.76)
(7, 89.54)
(8, 91.94)
(9, 93.88)
(10, 95.89)
(11, 96.74)
(12, 97.98)
(13, 98.63)
(14, 98.84)
(15, 99.27)
(16, 99.59)
(17, 99.73)
(18, 99.82)
(19, 99.97)
(20, 100)
(21, 100)
(22, 100)
(23, 100)
(24, 100)
(25, 100)
(26, 100)
(27, 100)
(28, 100)
(29, 100)
(30, 100)
(31, 100)
(32, 100)
(33, 100)
(34, 100)
(35, 100)
(36, 100)
(37, 100)
(38, 100)
(39, 100)
(40, 100)
		};
		\addlegendentry{$x=1 - Sim$}
		
		\addplot[dashed, black] plot coordinates {
			(1, 0)
(2, 0)
(3, 0)
(4, 0)
(5, 0)
(6, 0.18)
(7, 4.8)
(8, 8.76)
(9, 15.45)
(10, 22.82)
(11, 30.82)
(12, 39.52)
(13, 47.5)
(14, 54.22)
(15, 62.57)
(16, 69.95)
(17, 75.5)
(18, 80.22)
(19, 84.37)
(20, 88.92)
(21, 90.22)
(22, 92.35)
(23, 95.3)
(24, 96.3)
(25, 97.55)
(26, 98.22)
(27, 99.05)
(28, 98.92)
(29, 99.45)
(30, 99.6)
(31, 99.82)
(32, 99.87)
(33, 99.92)
(34, 99.9)
(35, 99.97)
(36, 100)
(37, 100)
(38, 100)
(39, 100)
(40, 100)		
		};
		\addlegendentry{$x=5 - Model$}
		\addplot[only marks, mark=triangle, black] plot coordinates {
			(1, 0)
(2, 0)
(3, 0)
(4, 0)
(5, 0.04)
(6, 0.58)
(7, 3.76)
(8, 6.94)
(9, 13.25)
(10, 21.82)
(11, 28.82)
(12, 34.52)
(13, 45.5)
(14, 51.22)
(15, 59.57)
(16, 65.95)
(17, 70.32)
(18, 76.22)
(19, 81.37)
(20, 85.92)
(21, 87.22)
(22, 89.35)
(23, 92.3)
(24, 94.3)
(25, 95.55)
(26, 96.22)
(27, 96.95)
(28, 97.72)
(29, 98.25)
(30, 98.6)
(31, 99.2)
(32, 99.47)
(33, 99.62)
(34, 99.81)
(35, 99.94)
(36, 100)
(37, 100)
(38, 100)
(39, 100)
(40, 100)	
		};
		\addlegendentry{$x=5 - Sim$}
		
		\addplot[dotted,black] plot coordinates {
			(1, 0)
			(2, 0)
			(3, 0)
			(4, 0)
			(5, 0)
			(6, 0)
			(7, 0)
			(8, 0)
			(9, 0)
			(10, 0)
			(11, 0.075)
			(12, 0.425)
			(13, 1.05)
			(14, 2.575)
			(15, 4.225)
			(16, 7.45)
			(17, 12.075)
			(18, 17.15)
			(19, 24.92)
			(20, 30.15)
			(21, 36.77)
			(22, 47.13)
			(23, 54.35)
			(24, 61.47)
			(25, 67.85)
			(26, 74.65)
			(27, 80.575)
			(28, 84.17)
			(29, 88.27)
			(30, 91.22)
			(31, 93.6)
			(32, 95.82)
			(33, 97.37)
			(34, 98.4)
			(35, 99.02)
			(36, 99.42)
			(37, 99.07)
			(38, 99.82)
			(39, 99.87)
			(40, 100)	
		};
		\addlegendentry{$x=10 - Model$}
		\addplot[ only marks, mark=o,black] plot coordinates {
			(1, 0)
(2, 0)
(3, 0)
(4, 0)
(5, 0)
(6, 0)
(7, 0)
(8, 0)
(9, 0)
(10, 0.02)
(11, 0.146)
(12, 0.525)
(13, 1.25)
(14, 2.465)
(15, 3.825)
(16, 6.45)
(17, 10.575)
(18, 16.55)
(19, 19.72)
(20, 26.53)
(21, 33.77)
(22, 39.13)
(23, 47.35)
(24, 56.47)
(25, 63.85)
(26, 68.65)
(27, 74.575)
(28, 79.17)
(29, 82.27)
(30, 85.22)
(31, 88.6)
(32, 90.82)
(33, 92.37)
(34, 94.4)
(35, 96.02)
(36, 97.42)
(37, 98.07)
(38, 98.82)
(39, 99.17)
(40, 99.57)	
		};
		\addlegendentry{$x=10 - Sim$}
	\addplot[dashdotted,black] plot coordinates {
	(1, 0)
	(2, 0)
	(3, 0)
	(4, 0)
	(5, 0)
	(6, 0)
	(7, 0)
	(8, 0)
	(9, 0)
	(10, 0)
	(11, 0)
	(12, 0)
	(13, 0)
	(14, 0)
	(15, 0)
	(16, 0)
	(17, 0)
	(18, 0)
	(19, 0)
	(20, 0)
	(21, 0)
	(22, 0)
	(23, 0)
	(24, 0)
	(25, 0)
	(26, 0)
	(27, 0)
	(28, 0)
	(29, 0)
	(30, 0)
	(31, 0)
	(32, 0)
	(33, 0)
	(34, 0)
	(35, 0)
	(36, 0)
	(37, 0)
	(38, 0)
	(39, 1)
	(40, 2)	
};
\addlegendentry{$x=32 - Model$}
\addplot[ only marks, mark=star,black] plot coordinates {
	(1, 0)
(2, 0)
(3, 0)
(4, 0)
(5, 0)
(6, 0)
(7, 0)
(8, 0)
(9, 0)
(10, 0)
(11, 0)
(12, 0)
(13, 0)
(14, 0)
(15, 0)
(16, 0)
(17, 0)
(18, 0)
(19, 0)
(20, 0)
(21, 0)
(22, 0)
(23, 0)
(24, 0)
(25, 0)
(26, 0)
(27, 0)
(28, 0)
(29, 0)
(30, 0)
(31, 0)
(32, 0)
(33, 0)
(34, 0)
(35, 0)
(36, 0)
(37, 0)
(38, 0)
(39, 1)
(40, 1)	
};
\addlegendentry{$x=32 - Sim$}
		\end{axis}
		\end{tikzpicture}}
	\qquad
	\subfloat[$q=2, w=4$]{\begin{tikzpicture}
		[scale= 0.55]
		\begin{axis}[
		xlabel= $m$,
		ylabel=$Probailibity\; of\; Decoding\; x\; Packets$,
		xlabel style={font=\large},
		grid=major,
		legend pos= outer north east]
		
		\addplot[black] plot coordinates {
			(1, 7.87)
			(2, 14.15)
			(3, 20.52)
			(4, 26.75)
			(5, 32.1)
			(6, 38.25)
			(7, 40.87)
			(8, 45.52)
			(9, 52.35)
			(10, 53.52)
			(11, 59.22)
			(12, 62.3)
			(13, 64.9)
			(14, 67.95)
			(15, 70.8)
			(16, 74.15)
			(17, 74.77)
			(18, 77.85)
			(19, 80.15)
			(20, 81.55)
			(21, 84.92)
			(22, 86.92)
			(23, 89.32)
			(24, 90.82)
			(25, 92.07)
			(26, 94.75)
			(27, 96.77)
			(28, 98.8)
			(29, 99.22)
			(30, 99.77)
			(31, 99.92)
			(32, 99.97)
			(33, 100)
			(34, 100)
			(35, 100)
			(36, 100)
			(37, 100)
			(38, 100)
			(39, 100)
			(40, 100)		
		};
		\addlegendentry{$x=1 - Model$}
		\addplot[only marks,mark=square,black] plot coordinates {
			(1, 7.59)
			(2, 12.83)
			(3, 18.35)
			(4, 23.75)
			(5, 29.1)
			(6, 34.25)
			(7, 38.87)
			(8, 43.52)
			(9, 49.35)
			(10, 52.52)
			(11, 56.22)
			(12, 59.3)
			(13, 63.9)
			(14, 67.82)
			(15, 70.47)
			(16, 73.95)
			(17, 75.77)
			(18, 77.58)
			(19, 80.45)
			(20, 82.55)
			(21, 84.62)
			(22, 86.52)
			(23, 88.72)
			(24, 90.82)
			(25, 92.47)
			(26, 94.57)
			(27, 96.73)
			(28, 98.78)
			(29, 99.73)
			(30, 99.89)
			(31, 100)
			(32, 100)
			(33, 100)
			(34, 100)
			(35, 100)
			(36, 100)
			(37, 100)
			(38, 100)
			(39, 100)
			(40, 100)
		};
		\addlegendentry{$x=1-Sim$}
		
		\addplot[dashed, black] plot coordinates {
			(1, 0)
			(2, 0)
			(3, 0)
			(4, 0)
			(5, 0)
			(6, 0)
			(7, 0.075)
			(8, 0.1)
			(9, 0.125)
			(10, 0.3)
			(11, 0.575)
			(12, 0.825)
			(13, 1.225)
			(14, 1.7)
			(15, 2.625)
			(16, 3.35)
			(17, 3.95)
			(18, 5.4)
			(19, 7.35)
			(20, 9.9)
			(21, 11.17)
			(22, 16)
			(23, 18.62)
			(24, 24.62)
			(25, 29.07)
			(26, 38.4)
			(27, 50.25)
			(28, 63.72)
			(29, 79.37)
			(30, 85.25)
			(31, 93.77)
			(32, 98.02)
			(33, 99.75)
			(34, 99.8)
			(35, 100)
			(36, 100)
			(37, 100)
			(38, 100)
			(39, 100)
			(40, 100)		
		};
		\addlegendentry{$x=5-Model$}
		\addplot[only marks, mark=triangle, black] plot coordinates {
			(1, 0)
			(2, 0)
			(3, 0)
			(4, 0)
			(5, 0)
			(6, 0.02)
			(7, 0.25)
			(8, 0.4)
			(9, 0.825)
			(10, 1)
			(11, 1.375)
			(12, 2.825)
			(13, 2.725)
			(14, 4.7)
			(15, 5.625)
			(16, 6.35)
			(17, 6.95)
			(18, 8.4)
			(19, 9.35)
			(20, 10.9)
			(21, 14.17)
			(22, 16)
			(23, 18.62)
			(24, 22.62)
			(25, 26.07)
			(26, 38.4)
			(27, 51.25)
			(28, 66.72)
			(29, 77.37)
			(30, 82.25)
			(31, 88.77)
			(32, 94.02)
			(33, 97.75)
			(34, 98.8)
			(35, 99.4)
			(36, 100)
			(37, 100)
			(38, 100)
			(39, 100)
			(40, 100)	
		};
		\addlegendentry{$x=5-Sim$}
		
		\addplot[dotted,black] plot coordinates {
			(1, 0)
			(2, 0)
			(3, 0)
			(4, 0)
			(5, 0)
			(6, 0)
			(7, 0)
			(8, 0)
			(9, 0)
			(10, 0)
			(11, 0)
			(12, 0)
			(13, 0)
			(14, 0)
			(15, 0)
			(16, 0)
			(17, 0)
			(18, 0.075)
			(19, 0.125)
			(20, 0.375)
			(21, 0.425)
			(22, 1.05)
			(23, 1.375)
			(24, 1.65)
			(25, 3.525)
			(26, 6.4)
			(27, 11.37)
			(28, 20.52)
			(29, 35.37)
			(30, 55.15)
			(31, 76.27)
			(32, 89.4)
			(33, 95.65)
			(34, 98.2)
			(35, 99.55)
			(36, 99.8)
			(37, 99.87)
			(38, 99.97)
			(39, 100)
			(40, 100)	
		};
		\addlegendentry{$x=10-Model$}
		\addplot[ only marks, mark=o,black] plot coordinates {
						(1, 0)
			(2, 0)
			(3, 0)
			(4, 0)
			(5, 0)
			(6, 0)
			(7, 0)
			(8, 0)
			(9, 0)
			(10, 0)
			(11, 0)
			(12, 0)
			(13, 0)
			(14, 0)
			(15, 0)
			(16, 0.01)
			(17, 0.03)
			(18, 0.065)
			(19, 0.105)
			(20, 0.295)
			(21, 0.375)
			(22, 0.892)
			(23, 1.275)
			(24, 1.5)
			(25, 2.525)
			(26, 5.4)
			(27, 10.37)
			(28, 18.52)
			(29, 31.37)
			(30, 49.15)
			(31, 69.27)
			(32, 84.4)
			(33, 92.65)
			(34, 95.2)
			(35, 98.55)
			(36, 99.32)
			(37, 99.68)
			(38, 100)
			(39, 100)
			(40, 100)
		};
		\addlegendentry{$x=10-Sim$}
			\addplot[dashdotted,black] plot coordinates {
			(1, 0)
			(2, 0)
			(3, 0)
			(4, 0)
			(5, 0)
			(6, 0)
			(7, 0)
			(8, 0)
			(9, 0)
			(10, 0)
			(11, 0)
			(12, 0)
			(13, 0)
			(14, 0)
			(15, 0)
			(16, 0)
			(17, 0)
			(18, 0)
			(19, 0)
			(20, 0)
			(21, 0)
			(22, 0)
			(23, 0)
			(24, 0)
			(25, 0)
			(26, 0)
			(27, 0)
			(28, 0)
			(29, 0)
			(30, 0)
			(31, 0)
			(32, 1.54)
(33, 6.82)
(34, 10.94)
(35, 27.96)
(36, 46.82)
(37, 73.18)
(38, 79.33)
(39, 82.82)
(40, 84.32)		
		};
		\addlegendentry{$x=32 - Model$}
		\addplot[ only marks, mark=star,black] plot coordinates {
			(1, 0)
			(2, 0)
			(3, 0)
			(4, 0)
			(5, 0)
			(6, 0)
			(7, 0)
			(8, 0)
			(9, 0)
			(10, 0)
			(11, 0)
			(12, 0)
			(13, 0)
			(14, 0)
			(15, 0)
			(16, 0)
			(17, 0)
			(18, 0)
			(19, 0)
			(20, 0)
			(21, 0)
			(22, 0)
			(23, 0)
			(24, 0)
			(25, 0)
			(26, 0)
			(27, 0)
			(28, 0)
			(29, 0)
			(30, 0)
			(31, 0)
			(32, 1.14)
(33, 6.62)
(34, 10.74)
(35, 27.66)
(36, 46.32)
(37, 72.88)
(38, 78.45)
(39, 81.35)
(40, 84.05)		
		};
		\addlegendentry{$x=32 - Sim$}
		\end{axis}
		\end{tikzpicture}}
	\\
	\subfloat[$q=2^{4}, w=2$]{\begin{tikzpicture}
		[scale= 0.55]
		\begin{axis}[
		xlabel= $m$,
		ylabel=$Probailibity\; of\; Decoding\; x\; Packets$,
		xlabel style={font=\large},
		grid=major,
		legend pos= outer north east]
		
		\addplot[black] plot coordinates {
			(1, 29.4)
			(2, 48.95)
			(3, 62.75)
			(4, 74.2)
			(5, 80.65)
			(6, 86.62)
			(7, 89.82)
			(8, 93.37)
			(9, 95.1)
			(10, 96.42)
			(11, 97.42)
			(12, 98.1)
			(13, 99.05)
			(14, 99.3)
			(15, 99.35)
			(16, 99.65)
			(17, 99.75)
			(18, 99.85)
			(19, 99.87)
			(20, 99.87)
			(21, 99.92)
			(22, 99.92)
			(23, 99.95)
			(24, 99.87)
			(25, 100)
			(26, 100)
			(27, 100)
			(28, 100)
			(29, 100)
			(30, 100)
			(31, 100)
			(32, 100)
			(33, 100)
			(34, 100)
			(35, 100)
			(36, 100)
			(37, 100)
			(38, 100)
			(39, 100)
			(40, 100)		
		};
		\addlegendentry{$x=1 - Model$}
		\addplot[only marks,mark=square,black] plot coordinates {
			(1, 27.4)
			(2, 44.95)
			(3, 60.75)
			(4, 72.2)
			(5, 79.65)
			(6, 83.62)
			(7, 86.82)
			(8, 90.37)
			(9, 92.1)
			(10, 94.42)
			(11, 96.42)
			(12, 97.1)
			(13, 97.75)
			(14, 98.3)
			(15, 98.85)
			(16, 99.05)
			(17, 99.45)
			(18, 99.75)
			(19, 99.85)
			(20, 99.95)
			(21, 100)
			(22, 100)
			(23, 100)
			(24, 100)
			(25, 100)
			(26, 100)
			(27, 100)
			(28, 100)
			(29, 100)
			(30, 100)
			(31, 100)
			(32, 100)
			(33, 100)
			(34, 100)
			(35, 100)
			(36, 100)
			(37, 100)
			(38, 100)
			(39, 100)
			(40, 100)	
		};
		\addlegendentry{$x=1-Sim$}
		
		\addplot[dashed, black] plot coordinates {
			(1, 0)
			(2, 0)
			(3, 0)
			(4, 0)
			(5, 0.225)
			(6, 1.975)
			(7, 4.425)
			(8, 8.9)
			(9, 14.67)
			(10, 22.67)
			(11, 30.47)
			(12, 36.7)
			(13, 47.82)
			(14, 53.92)
			(15, 62.25)
			(16, 69.22)
			(17, 74.1)
			(18, 78.72)
			(19, 82.37)
			(20, 85.42)
			(21, 90.32)
			(22, 91.6)
			(23, 93.45)
			(24, 94.7)
			(25, 95.97)
			(26, 97.4)
			(27, 97.87)
			(28, 98.8)
			(29, 99.07)
			(30, 99.22)
			(31, 99.45)
			(32, 99.62)
			(33, 99.8)
			(34, 99.85)
			(35, 99.97)
			(36, 99.85)
			(37, 99.92)
			(38, 100)
			(39, 100)
			(40, 100)		
		};
		\addlegendentry{$x=5-Model$}
		\addplot[only marks, mark=triangle, black] plot coordinates {
		(1, 0)
		(2, 0)
		(3, 0)
		(4, 0)
		(5, 0)
		(6, 1.022)
		(7, 3.425)
		(8, 7.9)
		(9, 13.67)
		(10, 19.67)
		(11, 28.47)
		(12, 35.7)
		(13, 45.82)
		(14, 52.92)
		(15, 59.25)
		(16, 65.22)
		(17, 71.1)
		(18, 76.72)
		(19, 81.37)
		(20, 84.42)
		(21, 89.32)
		(22, 90.6)
		(23, 92.45)
		(24, 93.7)
		(25, 94.97)
		(26, 95.94)
		(27, 96.87)
		(28, 97.8)
		(29, 98.87)
		(30, 99.42)
		(31, 99.65)
		(32, 99.82)
		(33, 99.9)
		(34, 99.95)
		(35, 100)
		(36, 100)
		(37, 100)
		(38, 100)
		(39, 100)
		(40, 100)	
		};
		\addlegendentry{$x=5-Sim$}
		
		\addplot[dotted,black] plot coordinates {
			(1, 0)
			(2, 0)
			(3, 0)
			(4, 0)
			(5, 0)
			(6, 0)
			(7, 0)
			(8, 0)
			(9, 0)
			(10, 0.025)
			(11, 0.05)
			(12, 0.475)
			(13, 1.375)
			(14, 2.35)
			(15, 4.225)
			(16, 7.9)
			(17, 12.075)
			(18, 16.875)
			(19, 23.7)
			(20, 29.175)
			(21, 37.82)
			(22, 35.65)
			(23, 51.1)
			(24, 57.92)
			(25, 67.32)
			(26, 72.97)
			(27, 76.87)
			(28, 82.62)
			(29, 86.9)
			(30, 89.4)
			(31, 92.67)
			(32, 94.65)
			(33, 96.27)
			(34, 97.7)
			(35, 98.27)
			(36, 98.47)
			(37, 99.1)
			(38, 99.52)
			(39, 99.67)
			(40, 99.85)	
		};
		\addlegendentry{$x=10-Model$}
		\addplot[ only marks, mark=o,black] plot coordinates {
					(1, 0)
		(2, 0)
		(3, 0)
		(4, 0)
		(5, 0)
		(6, 0)
		(7, 0)
		(8, 0)
		(9, 0)
		(10, 0)
		(11, 0.15)
		(12, 0.675)
		(13, 1.575)
		(14, 3.15)
		(15, 3.725)
		(16, 6.9)
		(17, 11.075)
		(18, 17.875)
		(19, 24.7)
		(20, 30.175)
		(21, 36.82)
		(22, 42.65)
		(23, 48.1)
		(24, 54.92)
		(25, 63.32)
		(26, 69.97)
		(27, 73.87)
		(28, 79.62)
		(29, 83.9)
		(30, 86.4)
		(31, 89.67)
		(32, 92.65)
		(33, 94.27)
		(34, 96.7)
		(35, 97.72)
		(36, 98.67)
		(37, 99.4)
		(38, 99.72)
		(39, 99.97)
		(40, 100)	
		};
		\addlegendentry{$x=10-Sim$}
			\addplot[dashdotted,black] plot coordinates {
			(1, 0)
			(2, 0)
			(3, 0)
			(4, 0)
			(5, 0)
			(6, 0)
			(7, 0)
			(8, 0)
			(9, 0)
			(10, 0)
			(11, 0)
			(12, 0)
			(13, 0)
			(14, 0)
			(15, 0)
			(16, 0)
			(17, 0)
			(18, 0)
			(19, 0)
			(20, 0)
			(21, 0)
			(22, 0)
			(23, 0)
			(24, 0)
			(25, 0)
			(26, 0)
			(27, 0)
			(28, 0)
			(29, 0)
			(30, 0)
			(31, 0)
			(32, 0)
			(33, 0)
			(34, 0)
			(35, 0)
			(36, 0)
			(37, 0)
			(38, 0)
			(39, 1)
			(40, 2)	
		};
		\addlegendentry{$x=32 - Model$}
		\addplot[ only marks, mark=star,black] plot coordinates {
			(1, 0)
			(2, 0)
			(3, 0)
			(4, 0)
			(5, 0)
			(6, 0)
			(7, 0)
			(8, 0)
			(9, 0)
			(10, 0)
			(11, 0)
			(12, 0)
			(13, 0)
			(14, 0)
			(15, 0)
			(16, 0)
			(17, 0)
			(18, 0)
			(19, 0)
			(20, 0)
			(21, 0)
			(22, 0)
			(23, 0)
			(24, 0)
			(25, 0)
			(26, 0)
			(27, 0)
			(28, 0)
			(29, 0)
			(30, 0)
			(31, 0)
			(32, 0)
			(33, 0)
			(34, 0)
			(35, 0)
			(36, 0)
			(37, 0)
			(38, 0)
			(39, 1)
			(40, 1)	
		};
		\addlegendentry{$x=32 - Sim$}
		\end{axis}
		\end{tikzpicture}}
	\qquad
	\subfloat[$q=2^{4}, w=4$]{\begin{tikzpicture}
		[scale= 0.55]
		\begin{axis}[
		xlabel= $m$,
		ylabel=$Probailibity\; of\; Decoding\; x\; Packets$,
		xlabel style={font=\large},
		grid=major,
		legend pos= outer north east]
		
		\addplot[black] plot coordinates {
			(1, 6.96)
			(2, 13.99)
			(3, 20.08)
			(4, 26.05)
			(5, 31.32)
			(6, 36.6)
			(7, 41.15)
			(8, 45.49)
			(9, 49.42)
			(10, 53.5)
			(11, 56.6)
			(12, 59.79)
			(13, 62.88)
			(14, 65.69)
			(15, 68.09)
			(16, 70.36)
			(17, 72.44)
			(18, 74.64)
			(19, 76.75)
			(20, 78.74)
			(21, 80.64)
			(22, 81.89)
			(23, 83.2)
			(24, 84.61)
			(25, 85.64)
			(26, 86.81)
			(27, 87.85)
			(28, 89.16)
			(29, 90.85)
			(30, 93.6)
			(31, 95.3)
			(32, 98.1)
			(33, 99.8)
			(34, 99.98)
			(35, 100)
			(36, 100)
			(37, 100)
			(38, 100)
			(39, 100)
			(40, 100)		
		};
		\addlegendentry{$x=1 - Model$}
		\addplot[only marks,mark=square,black] plot coordinates {
		(1, 8.14)
		(2, 14.62)
		(3, 20.47)
		(4, 25.85)
		(5, 30.72)
		(6, 35.96)
		(7, 40.15)
		(8, 44.99)
		(9, 48.82)
		(10, 52.5)
		(11, 56.4)
		(12, 60.79)
		(13, 63.88)
		(14, 66.22)
		(15, 68.49)
		(16, 70.62)
		(17, 72.74)
		(18, 74.4)
		(19, 76.5)
		(20, 78.47)
		(21, 80.46)
		(22, 82.39)
		(23, 83.7)
		(24, 84.91)
		(25, 86.14)
		(26, 87.41)
		(27, 88.15)
		(28, 89.26)
		(29, 90.45)
		(30, 91.76)
		(31, 93.13)
		(32, 95.28)
		(33, 97.8)
		(34, 99.06)
		(35, 100)
		(36, 100)
		(37, 100)
		(38, 100)
		(39, 100)
		(40, 100)		
		};
		\addlegendentry{$x=1-Sim$}
		
		\addplot[dashed, black] plot coordinates {
			(1, 0)
			(2, 0)
			(3, 0)
			(4, 0)
			(5, 0)
			(6, 0.01)
			(7, 0.01)
			(8, 0.02)
			(9, 0.09)
			(10, 0.22)
			(11, 0.4)
			(12, 0.7)
			(13, 1.06)
			(14, 1.46)
			(15, 1.93)
			(16, 2.72)
			(17, 3.38)
			(18, 4.28)
			(19, 5.42)
			(20, 6.66)
			(21, 7.95)
			(22, 9.42)
			(23, 11.01)
			(24, 12.83)
			(25, 14.76)
			(26, 16.93)
			(27, 19.71)
			(28, 23.26)
			(29, 29.13)
			(30, 40.82)
			(31, 60.57)
			(32, 83.72)
			(33, 95.26)
			(34, 98.94)
			(35, 99.89)
			(36, 99.99)
			(37, 100)
			(38, 100)
			(39, 100)
			(40, 100)		
		};
		\addlegendentry{$x=5-Model$}
		\addplot[only marks, mark=triangle, black] plot coordinates {
						(1, 0)
			(2, 0)
			(3, 0)
			(4, 0)
			(5, 0)
			(6, 0)
			(7, 0.04)
			(8, 0.62)
			(9, 0.83)
			(10, 0.94)
			(11, 1.4)
			(12, 1.7)
			(13, 2.36)
			(14, 2.6)
			(15, 2.93)
			(16, 3.22)
			(17, 3.68)
			(18, 3.98)
			(19, 4.42)
			(20, 5.66)
			(21, 6.95)
			(22, 8.12)
			(23, 10.01)
			(24, 12.33)
			(25, 14.26)
			(26, 16.38)
			(27, 18.51)
			(28, 22.03)
			(29, 27.13)
			(30, 38.82)
			(31, 56.57)
			(32, 79.72)
			(33, 89.26)
			(34, 96.94)
			(35, 97.89)
			(36, 98.94)
			(37, 99.63)
			(38, 100)
			(39, 100)
			(40, 100)
		};
		\addlegendentry{$x=5-Sim$}
		
		\addplot[dotted,black] plot coordinates {
			(1, 0)
			(2, 0)
			(3, 0)
			(4, 0)
			(5, 0)
			(6, 0)
			(7, 0)
			(8, 0)
			(9, 0)
			(10, 0)
			(11, 0)
			(12, 0)
			(13, 0)
			(14, 0)
			(15, 0.01)
			(16, 0.02)
			(17, 0.03)
			(18, 0.09)
			(19, 0.17)
			(20, 0.27)
			(21, 0.41)
			(22, 0.53)
			(23, 0.74)
			(24, 1.08)
			(25, 1.44)
			(26, 2.11)
			(27, 3)
			(28, 4.58)
			(29, 8.15)
			(30, 17.1)
			(31, 36.28)
			(32, 71.34)
			(33, 90.98)
			(34, 97.97)
			(35, 99.58)
			(36, 99.95)
			(37, 100)
			(38, 100)
			(39, 100)
			(40, 100)	
		};
		\addlegendentry{$x=10-Model$}
		\addplot[ only marks, mark=o,black] plot coordinates {
						(1, 0)
			(2, 0)
			(3, 0)
			(4, 0)
			(5, 0)
			(6, 0)
			(7, 0)
			(8, 0)
			(9, 0)
			(10, 0)
			(11, 0)
			(12, 0)
			(13, 0)
			(14, 0)
			(15, 0)
			(16, 0.05)
			(17, 0.09)
			(18, 0.29)
			(19, 0.47)
			(20, 0.67)
			(21, 0.81)
			(22, 1.03)
			(23, 1.74)
			(24, 2.28)
			(25, 3.44)
			(26, 3.71)
			(27, 4)
			(28, 5.58)
			(29, 7.15)
			(30, 16.1)
			(31, 29.28)
			(32, 61.34)
			(33, 84.98)
			(34, 91.97)
			(35, 94.58)
			(36, 97.95)
			(37, 100)
			(38, 100)
			(39, 100)
		};
		\addlegendentry{$x=10-Sim$}
			\addplot[dashdotted,black] plot coordinates {
			(1, 0)
			(2, 0)
			(3, 0)
			(4, 0)
			(5, 0)
			(6, 0)
			(7, 0)
			(8, 0)
			(9, 0)
			(10, 0)
			(11, 0)
			(12, 0)
			(13, 0)
			(14, 0)
			(15, 0)
			(16, 0)
			(17, 0)
			(18, 0)
			(19, 0)
			(20, 0)
			(21, 0)
			(22, 0)
			(23, 0)
			(24, 0)
			(25, 0)
			(26, 0)
			(27, 0)
			(28, 0)
			(29, 0)
			(30, 0)
			(31, 0)
			(32, 1.34)
(33, 8.62)
(34, 15.84)
(35, 37.96)
(36, 59.82)
(37, 79.18)
(38, 86.63)
(39, 88.82)
(40, 92.32)		
		};
		\addlegendentry{$x=32 - Model$}
		\addplot[ only marks, mark=star,black] plot coordinates {
			(1, 0)
			(2, 0)
			(3, 0)
			(4, 0)
			(5, 0)
			(6, 0)
			(7, 0)
			(8, 0)
			(9, 0)
			(10, 0)
			(11, 0)
			(12, 0)
			(13, 0)
			(14, 0)
			(15, 0)
			(16, 0)
			(17, 0)
			(18, 0)
			(19, 0)
			(20, 0)
			(21, 0)
			(22, 0)
			(23, 0)
			(24, 0)
			(25, 0)
			(26, 0)
			(27, 0)
			(28, 0)
			(29, 0)
			(30, 0)
			(31, 0)
			(32, 1.34)
(33, 6.42)
(34, 13.97)
(35, 32.76)
(36, 57.82)
(37, 77.18)
(38, 84.33)
(39, 86.82)
(40, 91.32)	
		};
		\addlegendentry{$x=32 - Sim$}
		\end{axis}
		
		\end{tikzpicture}}
		\\
	\subfloat[$q=2^{8}, w=2$]{\begin{tikzpicture}
		[scale= 0.55]
		\begin{axis}[
		xlabel= $m$,
		ylabel=$Probailibity\; of\; Decoding\; x\; Packets$,
		xlabel style={font=\large},
		grid=major,
		legend pos= outer north east]
		
		\addplot[black] plot coordinates {
			(1, 27.4)
			(2, 45.95)
			(3, 60.75)
			(4, 73.2)
			(5, 79.65)
			(6, 85.62)
			(7, 89.82)
			(8, 93.27)
			(9, 94.91)
			(10, 96.12)
			(11, 96.82)
			(12, 97.41)
			(13, 98.25)
			(14, 98.83)
			(15, 99.35)
			(16, 99.65)
			(17, 99.75)
			(18, 99.85)
			(19, 99.87)
			(20, 99.87)
			(21, 99.92)
			(22, 99.92)
			(23, 99.95)
			(24, 99.87)
			(25, 100)
			(26, 100)
			(27, 100)
			(28, 100)
			(29, 100)
			(30, 100)
			(31, 100)
			(32, 100)
			(33, 100)
			(34, 100)
			(35, 100)
			(36, 100)
			(37, 100)
			(38, 100)
			(39, 100)
			(40, 100)		
		};
		\addlegendentry{$x=1 - Model$}
		\addplot[only marks,mark=square,black] plot coordinates {
			(1, 25.4)
			(2, 42.95)
			(3, 59.75)
			(4, 72.2)
			(5, 79.65)
			(6, 83.62)
			(7, 86.82)
			(8, 90.37)
			(9, 92.1)
			(10, 94.42)
			(11, 96.42)
			(12, 97.1)
			(13, 97.75)
			(14, 98.3)
			(15, 98.85)
			(16, 99.05)
			(17, 99.45)
			(18, 99.75)
			(19, 99.85)
			(20, 99.95)
			(21, 100)
			(22, 100)
			(23, 100)
			(24, 100)
			(25, 100)
			(26, 100)
			(27, 100)
			(28, 100)
			(29, 100)
			(30, 100)
			(31, 100)
			(32, 100)
			(33, 100)
			(34, 100)
			(35, 100)
			(36, 100)
			(37, 100)
			(38, 100)
			(39, 100)
			(40, 100)	
		};
		\addlegendentry{$x=1-Sim$}
		
		\addplot[dashed, black] plot coordinates {
			(1, 0)
			(2, 0)
			(3, 0)
			(4, 0)
			(5, 0.165)
			(6, 1.675)
			(7, 4.125)
			(8, 8.6)
			(9, 13.67)
			(10, 21.67)
			(11, 28.47)
			(12, 34.7)
			(13, 43.82)
			(14, 52.92)
			(15, 61.25)
			(16, 65.22)
			(17, 73.1)
			(18, 77.72)
			(19, 81.37)
			(20, 83.42)
			(21, 88.32)
			(22, 90.6)
			(23, 93.15)
			(24, 94.7)
			(25, 95.97)
			(26, 97.4)
			(27, 97.87)
			(28, 98.8)
			(29, 99.07)
			(30, 99.22)
			(31, 99.45)
			(32, 99.62)
			(33, 99.8)
			(34, 99.85)
			(35, 99.97)
			(36, 99.85)
			(37, 99.92)
			(38, 100)
			(39, 100)
			(40, 100)		
		};
		\addlegendentry{$x=5-Model$}
		\addplot[only marks, mark=triangle, black] plot coordinates {
			(1, 0)
			(2, 0)
			(3, 0)
			(4, 0)
			(5, 0)
			(6, 0.822)
			(7, 3.125)
			(8, 7.6)
			(9, 13.37)
			(10, 19.47)
			(11, 28.07)
			(12, 35.3)
			(13, 45.42)
			(14, 52.62)
			(15, 59.05)
			(16, 65.12)
			(17, 70.8)
			(18, 76.72)
			(19, 81.07)
			(20, 84.22)
			(21, 89.02)
			(22, 89.8)
			(23, 91.45)
			(24, 91.7)
			(25, 93.47)
			(26, 95.34)
			(27, 96.47)
			(28, 97.5)
			(29, 97.87)
			(30, 99.12)
			(31, 99.25)
			(32, 99.62)
			(33, 99.84)
			(34, 99.88)
			(35, 99.92)
			(36, 100)
			(37, 100)
			(38, 100)
			(39, 100)
			(40, 100)	
		};
		\addlegendentry{$x=5-Sim$}
		
		\addplot[dotted,black] plot coordinates {
			(1, 0)
			(2, 0)
			(3, 0)
			(4, 0)
			(5, 0)
			(6, 0)
			(7, 0)
			(8, 0)
			(9, 0)
			(10, 0.0125)
			(11, 0.035)
			(12, 0.425)
			(13, 1.175)
			(14, 2.05)
			(15, 4.025)
			(16, 7.67)
			(17, 11.75)
			(18, 16.375)
			(19, 20.21)
			(20, 25.175)
			(21, 32.82)
			(22, 38.65)
			(23, 45.1)
			(24, 54.92)
			(25, 65.32)
			(26, 69.97)
			(27, 73.87)
			(28, 79.62)
			(29, 83.9)
			(30, 85.4)
			(31, 89.67)
			(32, 93.65)
			(33, 95.27)
			(34, 96.7)
			(35, 97.87)
			(36, 98.27)
			(37, 99.13)
			(38, 99.52)
			(39, 99.67)
			(40, 99.85)	
		};
		\addlegendentry{$x=10-Model$}
		\addplot[ only marks, mark=o,black] plot coordinates {
			(1, 0)
			(2, 0)
			(3, 0)
			(4, 0)
			(5, 0)
			(6, 0)
			(7, 0)
			(8, 0)
			(9, 0)
			(10, 0)
			(11, 0.05)
			(12, 0.475)
			(13, 1.275)
			(14, 3.15)
			(15, 3.525)
			(16, 7.37)
			(17, 11.25)
			(18, 15.975)
			(19, 20.1)
			(20, 23.175)
			(21, 32.52)
			(22, 38.15)
			(23, 44.7)
			(24, 52.92)
			(25, 64.32)
			(26, 68.97)
			(27, 72.87)
			(28, 77.62)
			(29, 82.4)
			(30, 84.4)
			(31, 87.67)
			(32, 91.65)
			(33, 94.27)
			(34, 95.7)
			(35, 96.87)
			(36, 97.77)
			(37, 98.93)
			(38, 99.2)
			(39, 99.57)
			(40, 99.93)	
		};
		\addlegendentry{$x=10-Sim$}
			\addplot[dashdotted,black] plot coordinates {
			(1, 0)
			(2, 0)
			(3, 0)
			(4, 0)
			(5, 0)
			(6, 0)
			(7, 0)
			(8, 0)
			(9, 0)
			(10, 0)
			(11, 0)
			(12, 0)
			(13, 0)
			(14, 0)
			(15, 0)
			(16, 0)
			(17, 0)
			(18, 0)
			(19, 0)
			(20, 0)
			(21, 0)
			(22, 0)
			(23, 0)
			(24, 0)
			(25, 0)
			(26, 0)
			(27, 0)
			(28, 0)
			(29, 0)
			(30, 0)
			(31, 0)
			(32, 0)
			(33, 0)
			(34, 0)
			(35, 0)
			(36, 0)
			(37, 0)
			(38, 0)
			(39, 1)
			(40, 2)	
		};
		\addlegendentry{$x=32 - Model$}
		\addplot[ only marks, mark=star,black] plot coordinates {
			(1, 0)
			(2, 0)
			(3, 0)
			(4, 0)
			(5, 0)
			(6, 0)
			(7, 0)
			(8, 0)
			(9, 0)
			(10, 0)
			(11, 0)
			(12, 0)
			(13, 0)
			(14, 0)
			(15, 0)
			(16, 0)
			(17, 0)
			(18, 0)
			(19, 0)
			(20, 0)
			(21, 0)
			(22, 0)
			(23, 0)
			(24, 0)
			(25, 0)
			(26, 0)
			(27, 0)
			(28, 0)
			(29, 0)
			(30, 0)
			(31, 0)
			(32, 0)
			(33, 0)
			(34, 0)
			(35, 0)
			(36, 0)
			(37, 0)
			(38, 0)
			(39, 1)
			(40, 1)	
		};
		\addlegendentry{$x=32 - Sim$}
		\end{axis}
		\end{tikzpicture}}
	\qquad
	\subfloat[$q=2^{8}, w=4$]{\begin{tikzpicture}
		[scale= 0.55]
		\begin{axis}[
		xlabel= $m$,
		ylabel=$Probailibity\; of\; Decoding\; x\; Packets$,
		xlabel style={font=\large},
		grid=major,
		legend pos= outer north east]
		
		\addplot[black] plot coordinates {
			(1, 5.46)
			(2, 12.29)
			(3, 18.08)
			(4, 24.65)
			(5, 30.12)
			(6, 35.03)
			(7, 40.35)
			(8, 44.69)
			(9, 48.82)
			(10, 52.5)
			(11, 55.8)
			(12, 59.09)
			(13, 62.18)
			(14, 64.9)
			(15, 67.9)
			(16, 69.86)
			(17, 71.84)
			(18, 73.94)
			(19, 76.25)
			(20, 78.4)
			(21, 79.94)
			(22, 80.9)
			(23, 81.8)
			(24, 83.51)
			(25, 84.4)
			(26, 85.1)
			(27, 86.75)
			(28, 88.1)
			(29, 89.75)
			(30, 91.6)
			(31, 93.43)
			(32, 97)
			(33, 98)
			(34, 98.8)
			(35, 99.2)
			(36, 99.6)
			(37, 99.8)
			(38, 100)
			(39, 100)
			(40, 100)		
		};
		\addlegendentry{$x=1 - Model$}
		\addplot[only marks,mark=square,black] plot coordinates {
		(1, 5.56)
		(2, 12.69)
		(3, 17.8)
		(4, 24.35)
		(5, 30.2)
		(6, 34.83)
		(7, 40.5)
		(8, 44.9)
		(9, 49.2)
		(10, 52.8)
		(11, 56.5)
		(12, 59.9)
		(13, 62.8)
		(14, 65.3)
		(15, 67.6)
		(16, 69.6)
		(17, 71.4)
		(18, 73.4)
		(19, 76.5)
		(20, 78.14)
		(21, 79.4)
		(22, 80.3)
		(23, 81.2)
		(24, 83.1)
		(25, 84.04)
		(26, 85.31)
		(27, 86.55)
		(28, 88.31)
		(29, 89.5)
		(30, 91.26)
		(31, 93.3)
		(32, 95.42)
		(33, 97.3)
		(34, 98.4)
		(35, 98.9)
		(36, 99.8)
		(37, 99.9)
		(38, 100)
		(39, 100)
		(40, 100)	
		};
		\addlegendentry{$x=1-Sim$}
		
		\addplot[dashed, black] plot coordinates {
			(1, 0)
(2, 0)
(3, 0)
(4, 0)
(5, 0)
(6, 0)
(7, 0.06)
(8, 0.42)
(9, 0.67)
(10, 0.94)
(11, 1.24)
(12, 1.47)
(13, 1.66)
(14, 1.96)
(15, 2.3)
(16, 2.52)
(17, 2.78)
(18, 2.86)
(19, 2.93)
(20, 4.46)
(21, 5.15)
(22, 5.52)
(23, 7.31)
(24, 9.35)
(25, 11.6)
(26, 13.58)
(27, 15.1)
(28, 18.3)
(29, 21.3)
(30, 30.2)
(31, 44.7)
(32, 61.2)
(33, 78.6)
(34, 86.4)
(35, 94.29)
(36, 96.4)
(37, 98.3)
(38, 99.1)
(39, 100)
(40, 100)		
		};
		\addlegendentry{$x=5-Model$}
		\addplot[only marks, mark=triangle, black] plot coordinates {
			(1, 0)
			(2, 0)
			(3, 0)
			(4, 0)
			(5, 0)
			(6, 0)
			(7, 0.04)
			(8, 0.32)
			(9, 0.63)
			(10, 0.84)
			(11, 1.14)
			(12, 1.37)
			(13, 1.76)
			(14, 2.06)
			(15, 2.43)
			(16, 2.62)
			(17, 2.8)
			(18, 2.98)
			(19, 3.12)
			(20, 4.26)
			(21, 4.95)
			(22, 5.32)
			(23, 7.41)
			(24, 9.53)
			(25, 11.26)
			(26, 13.38)
			(27, 15.51)
			(28, 18.03)
			(29, 21.13)
			(30, 30.82)
			(31, 44.57)
			(32, 61.72)
			(33, 78.26)
			(34, 86.94)
			(35, 93.89)
			(36, 96.94)
			(37, 98.63)
			(38, 99.81)
			(39, 100)
			(40, 100)
		};
		\addlegendentry{$x=5-Sim$}
		
		\addplot[dotted,black] plot coordinates {

(1, 0)
(2, 0)
(3, 0)
(4, 0)
(5, 0)
(6, 0)
(7, 0)
(8, 0)
(9, 0)
(10, 0)
(11, 0)
(12, 0)
(13, 0)
(14, 0)
(15, 0.01)
(16, 0.03)
(17, 0.06)
(18, 0.15)
(19, 0.26)
(20, 0.41)
(21, 0.64)
(22, 0.83)
(23, 1.04)
(24, 1.28)
(25, 2.34)
(26, 2.71)
(27, 3.16)
(28, 3.8)
(29, 4.25)
(30, 10.9)
(31, 23.6)
(32, 49.34)
(33, 68.38)
(34, 81.94)
(35, 89.58)
(36, 93.42)
(37, 96.89)
(38, 98.42)
(39, 99.4)
(40, 100)	
		};
		\addlegendentry{$x=10-Model$}
		\addplot[ only marks, mark=o,black] plot coordinates {
		(1, 0)
		(2, 0)
		(3, 0)
		(4, 0)
		(5, 0)
		(6, 0)
		(7, 0)
		(8, 0)
		(9, 0)
		(10, 0)
		(11, 0)
		(12, 0)
		(13, 0)
		(14, 0)
		(15, 0.02)
		(16, 0.05)
		(17, 0.07)
		(18, 0.18)
		(19, 0.29)
		(20, 0.47)
		(21, 0.71)
		(22, 0.89)
		(23, 1.14)
		(24, 1.38)
		(25, 2.54)
		(26, 2.81)
		(27, 3.26)
		(28, 3.9)
		(29, 4.35)
		(30, 11.9)
		(31, 24.6)
		(32, 47.94)
		(33, 67.88)
		(34, 80.84)
		(35, 88.58)
		(36, 92.62)
		(37, 96.19)
		(38, 97.92)
		(39, 99.1)
		(40, 99.8)
		};
		\addlegendentry{$x=10-Sim$}
			\addplot[dashdotted,black] plot coordinates {
			(1, 0)
			(2, 0)
			(3, 0)
			(4, 0)
			(5, 0)
			(6, 0)
			(7, 0)
			(8, 0)
			(9, 0)
			(10, 0)
			(11, 0)
			(12, 0)
			(13, 0)
			(14, 0)
			(15, 0)
			(16, 0)
			(17, 0)
			(18, 0)
			(19, 0)
			(20, 0)
			(21, 0)
			(22, 0)
			(23, 0)
			(24, 0)
			(25, 0)
			(26, 0)
			(27, 0)
			(28, 0)
			(29, 0)
			(30, 0)
			(31, 0)
			(32, 1.34)
			(33, 7.62)
			(34, 14.84)
			(35, 33.96)
			(36, 57.82)
			(37, 77.18)
			(38, 84.33)
			(39, 88.82)
			(40, 91.32)	
		};
		\addlegendentry{$x=32 - Model$}
		\addplot[ only marks, mark=star,black] plot coordinates {
			(1, 0)
			(2, 0)
			(3, 0)
			(4, 0)
			(5, 0)
			(6, 0)
			(7, 0)
			(8, 0)
			(9, 0)
			(10, 0)
			(11, 0)
			(12, 0)
			(13, 0)
			(14, 0)
			(15, 0)
			(16, 0)
			(17, 0)
			(18, 0)
			(19, 0)
			(20, 0)
			(21, 0)
			(22, 0)
			(23, 0)
			(24, 0)
			(25, 0)
			(26, 0)
			(27, 0)
			(28, 0)
			(29, 0)
			(30, 0)
			(31, 0)
			(32, 1.34)
(33, 7.62)
(34, 14.84)
(35, 33.96)
(36, 57.82)
(37, 77.18)
(38, 84.33)
(39, 86.82)
(40, 91.32)		
		};
		\addlegendentry{$x=32 - Sim$}
		\end{axis}
		\end{tikzpicture}}
	\caption[2]{Simulation and theoretical results for probability of decoding at least $x$ source packets for $x=1,5,10,32$. $q=2,2^{4},2^{8}$, $w=2,4$ and $n=32$ in fully random SNC scheme}
	\label{fig:1}
\end{figure*}

Fig.~\ref{fig:1} shows the results of model for probability of decoding at least $x$ source packets after $m$ transmissions as well as simulation results for $n=32$, $x = \{1,5,10,32 \}$, $q\in \{2, 2^{4}, 2^{8}\}$ and $w\in\{2, 4\}$, where $p=\frac{w}{n}$. The theoretical results of these figures are generated using~\eqref{thefinal}. The results show that our model can predict the probability of decoding at least $x$ source packets from $m$ received coded packets with the average MAPD of 6~\%. These figures show that in lower sparsity, the decoder can recover a fraction of source packets in lower number of transmissions, for example, the probability of recovering at least one of the source packets is higher when $w=2$ with respect to $w=4$. however, the probability of decoding the whole generation is lower for lower sparsity, showing that the total number of required transmissions to decode the whole generation increases by lowering sparsity.

\begin{figure*}[!t]
	\centering
	\subfloat[$n =32$]{\begin{tikzpicture}
		[scale= 0.6]
		\begin{axis}[
		xlabel= $m$,
		ylabel=$p$,
		xlabel style={font=\large},
		grid=major,
		legend pos= north west]
		
		\addplot[black] plot coordinates {
			(1, 0.0454)
			(2, 0.0567)
			(3, 0.0585)
			(4, 0.0624)
			(5, 0.0684)
			(6, 0.0704)
			(7, 0.0737)
			(8, 0.0792)
			(9, 0.0829)
			(10, 0.0844)
			(11, 0.0937)
			(12, 0.0977)
			(13, 0.1058)
			(14, 0.1084)
			(15, 0.1148)
			(16, 0.1225)
			(17, 0.1328)
			(18, 0.1362)
			(19, 0.1457)
			(20, 0.166)
			(21, 0.189)
			(22, 0.216)
			(23, 0.268)
			(24, 0.294)
			(25, 0.324)
			(26, 0.367)
			(27, 0.416)
			(28, 0.448)
			(29, 0.473)
			(30, 0.488)
			(31, 0.494)
			(32, 0.5)		
		};
		\addlegendentry{$q=2$}
		\addplot[dashed, black] plot coordinates {
			(1, 0.0424)
			(2, 0.0467)
			(3, 0.0525)
			(4, 0.0594)
			(5, 0.0684)
			(6, 0.0724)
			(7, 0.0747)
			(8, 0.0762)
			(9, 0.0794)
			(10, 0.0814)
			(11, 0.0837)
			(12, 0.0847)
			(13, 0.0858)
			(14, 0.0864)
			(15, 0.0888)
			(16, 0.0925)
			(17, 0.0928)
			(18, 0.0942)
			(19, 0.0957)
			(20, 0.106)
			(21, 0.119)
			(22, 0.126)
			(23, 0.168)
			(24, 0.194)
			(25, 0.224)
			(26, 0.267)
			(27, 0.356)
			(28, 0.428)
			(29, 0.473)
			(30, 0.488)
			(31, 0.494)
			(32, 0.5)		
		};
		\addlegendentry{$q=2^{4}$}
		\addplot[dotted, black] plot coordinates {
			(1, 0.0418)
			(2, 0.0447)
			(3, 0.0515)
			(4, 0.0574)
			(5, 0.0664)
			(6, 0.0684)
			(7, 0.0707)
			(8, 0.0742)
			(9, 0.0764)
			(10, 0.0794)
			(11, 0.0807)
			(12, 0.0823)
			(13, 0.0842)
			(14, 0.0855)
			(15, 0.0867)
			(16, 0.0904)
			(17, 0.0918)
			(18, 0.0922)
			(19, 0.0937)
			(20, 0.0961)
			(21, 0.0994)
			(22, 0.101)
			(23, 0.118)
			(24, 0.144)
			(25, 0.164)
			(26, 0.184)
			(27, 0.276)
			(28, 0.348)
			(29, 0.393)
			(30, 0.438)
			(31, 0.474)
			(32, 0.493)		
		};
		\addlegendentry{$q=2^{8}$}
		\end{axis}
		\end{tikzpicture}}
	\qquad
	\subfloat[$n=64$]{\begin{tikzpicture}
		[scale= 0.6]
		\begin{axis}[
		xlabel= $m$,
		ylabel=$p$,
		xlabel style={font=\large},
		grid=major,
		legend pos= north west]
		\addplot[black] plot coordinates {
			(1, 0.0144)
			(2, 0.0182)
			(3, 0.0187)
			(4, 0.0214)
			(5, 0.0245)
			(6, 0.0288)
			(7, 0.0314)
			(8, 0.0317)
			(9, 0.0344)
			(10, 0.0394)
			(11, 0.0414)
			(12, 0.0426)
			(13, 0.0447)
			(14, 0.0463)
			(15, 0.0482)
			(16, 0.0499)
			(17, 0.0514)
			(18, 0.0523)
			(19, 0.0534)
			(20, 0.0547)
			(21, 0.0557)
			(22, 0.0562)
			(23, 0.0567)
			(24, 0.0572)
			(25, 0.0598)
			(26, 0.0637)
			(27, 0.0676)
			(28, 0.0693)
			(29, 0.0724)
			(30, 0.0743)
			(31, 0.0785)
			(32, 0.0837)
			(33, 0.0881)
			(34, 0.0918)
			(35, 0.1001)
			(36, 0.1023)
			(37, 0.1087)
			(38, 0.1103)
			(39, 0.118)
			(40, 0.136)
			(41, 0.151)
			(42, 0.165)
			(43, 0.178)
			(44, 0.201)
			(45, 0.22)
			(46, 0.233)
			(47, 0.249)
			(48, 0.27)
			(49, 0.286)
			(50, 0.315)
			(51, 0.339)
			(52, 0.355)
			(53, 0.388)
			(54, 0.418)
			(55, 0.437)
			(56, 0.459)
			(57, 0.476)
			(58, 0.498)
			(59, 0.5)
			(60, 0.5)
			(61, 0.5)
			(62, 0.5)
			(63, 0.5)
			(64, 0.5)	
		};
		\addlegendentry{$q=2$}
		\addplot[dashed, black] plot coordinates {
			(1, 0.0156)
			(2, 0.0158)
			(3, 0.0163)
			(4, 0.0165)
			(5, 0.0166)
			(6, 0.0169)
			(7, 0.0173)
			(8, 0.0178)
			(9, 0.0181)
			(10, 0.0194)
			(11, 0.0203)
			(12, 0.0216)
			(13, 0.0224)
			(14, 0.0243)
			(15, 0.0267)
			(16, 0.0276)
			(17, 0.0294)
			(18, 0.0314)
			(19, 0.0332)
			(20, 0.0348)
			(21, 0.0359)
			(22, 0.0384)
			(23, 0.0418)
			(24, 0.0442)
			(25, 0.0456)
			(26, 0.0471)
			(27, 0.0486)
			(28, 0.0499)
			(29, 0.0518)
			(30, 0.0532)
			(31, 0.0547)
			(32, 0.0573)
			(33, 0.0604)
			(34, 0.0628)
			(35, 0.0642)
			(36, 0.0717)
			(37, 0.0747)
			(38, 0.0776)
			(39, 0.0819)
			(40, 0.0932)
			(41, 0.1047)
			(42, 0.1136)
			(43, 0.1251)
			(44, 0.1382)
			(45, 0.153)
			(46, 0.170)
			(47, 0.196)
			(48, 0.208)
			(49, 0.231)
			(50, 0.253)
			(51, 0.289)
			(52, 0.323)
			(53, 0.356)
			(54, 0.376)
			(55, 0.408)
			(56, 0.438)
			(57, 0.459)
			(58, 0.473)
			(59, 0.495)
			(60, 0.5)
			(61, 0.5)
			(62, 0.5)
			(63, 0.5)
			(64, 0.5)	
		};
		\addlegendentry{$q=2^{4}$}
		\addplot[dotted, black] plot coordinates {
			(1, 0.0088)
			(2, 0.0091)
			(3, 0.0098)
			(4, 0.0102)
			(5, 0.0104)
			(6, 0.0117)
			(7, 0.0127)
			(8, 0.0135)
			(9, 0.0137)
			(10, 0.0144)
			(11, 0.0153)
			(12, 0.0164)
			(13, 0.0175)
			(14, 0.0186)
			(15, 0.0197)
			(16, 0.0209)
			(17, 0.0214)
			(18, 0.0226)
			(19, 0.0237)
			(20, 0.024)
			(21, 0.025)
			(22, 0.0275)
			(23, 0.0299)
			(24, 0.032)
			(25, 0.0337)
			(26, 0.035)
			(27, 0.036)
			(28, 0.037)
			(29, 0.04)
			(30, 0.042)
			(31, 0.0438)
			(32, 0.0453)
			(33, 0.0494)
			(34, 0.0508)
			(35, 0.0522)
			(36, 0.0597)
			(37, 0.0627)
			(38, 0.0656)
			(39, 0.0701)
			(40, 0.0712)
			(41, 0.0787)
			(42, 0.0816)
			(43, 0.0861)
			(44, 0.0922)
			(45, 0.1102)
			(46, 0.122)
			(47, 0.135)
			(48, 0.148)
			(49, 0.159)
			(50, 0.173)
			(51, 0.207)
			(52, 0.263)
			(53, 0.295)
			(54, 0.333)
			(55, 0.377)
			(56, 0.398)
			(57, 0.419)
			(58, 0.433)
			(59, 0.452)
			(60, 0.478)
			(61, 0.489)
			(62, 0.499)
			(63, 0.5)
			(64, 0.5)	
		};
		\addlegendentry{$q=2^{8}$}
		\end{axis}
		\end{tikzpicture}}
	\caption[2]{Evolution of sparsity regarding the number of received coded packets for $n=32,64$ and $q=2,2^{4},2^{8}$}
	\label{fig:3}
\end{figure*}
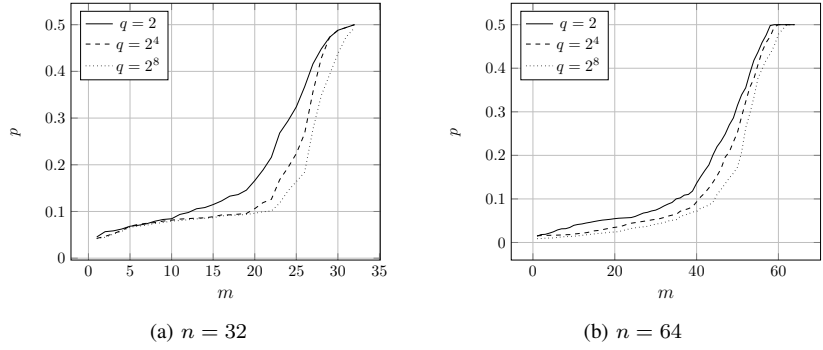

In order to test our tuning scheme, we used~\eqref{eq:theo2} to ~\eqref{Pprime} to calculate the best value for sparsity for each transmission. Fig.~\ref{fig:3} show the evolution of sparsity for $n\in \{32, 64\}$ and $q\in \{2,2^{4}\}$. As these figures show, the sparsity significantly increases when the number of received coded packets reaches $0.7\times n$. The impact of the Galois Field can also be clearly seen as the sparsity remains higher for higher Galois Field. This incident occurs due to the fact that in lower Galois Field, the source packets are decoded more quickly as can be seen in Fig.~\ref{fig:1}, hence, in order to increase the probability of decoding new source packets after each transmission, the sparsity tends to decrease more quickly.

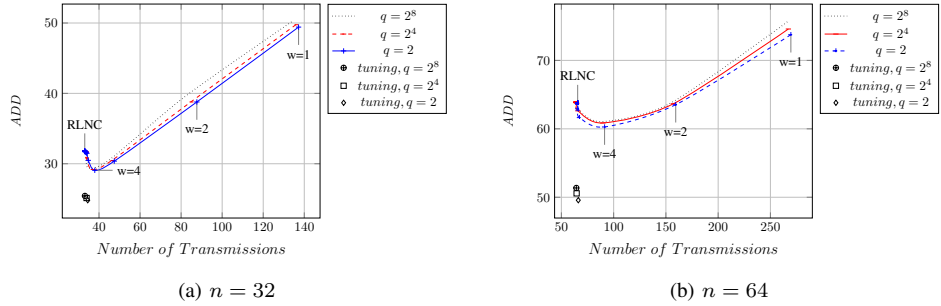
\begin{figure*}[!t]
	\centering
	\subfloat[$n =32$]{\begin{tikzpicture}
		[scale= 0.5]
		\begin{axis}[
		xlabel= $Number\; of\; Transmissions$,
		ylabel=$ADD$,
		xlabel style={font=\large},
		grid=major,
		legend pos= outer north east]
		
		\addplot[smooth,dotted, black] plot coordinates {
			(133.76, 50.16) 
			(82.31, 39.64) 
			(45.84, 30.73)
			(35.04, 29.34)
			(33.59, 31.03)
			(33.42, 31.72)
			(33.26, 31.75)
			(33.01, 31.78)
			(32.78, 31.8)
			(32.58, 31.81)
			(32.58, 31.82)
			(32.58, 31.84)
			(32.58, 31.84)
			(32.58, 31.84)
			(32.58, 31.84)
			(32.58, 31.84)
		};
		\addlegendentry{$q=2^{8}$}
		
		\addplot[smooth,mark=-,dashed, red] plot coordinates {
			(135.91, 49.76) 
			(84.73, 38.78) 
			(46.22, 30.46)
			(36.59, 29.18)
			(33.82, 30.83)
			(33.77, 31.66)
			(33.59, 31.72)
			(33.24, 31.76)
			(33.08, 31.77)
			(32.63, 31.78)
			(32.63, 31.81)
			(32.63, 31.83)
			(32.63, 31.84)
			(32.63, 31.84)
			(32.63, 31.84)
			(32.63, 31.84)
		};
		\addlegendentry{$q=2^{4}$}
		
		\addplot[smooth,mark=+,blue] plot coordinates {
					( 137.22, 49.42) 
			(87.66, 38.78) 
			(47.44, 30.37)
			(37.94, 29.07)
			(34.79, 30.48)
			(34.12, 31.48)
			(33.86, 31.56)
			(33.73, 31.66)
			(33.48, 31.69)
			(33.26, 31.71)
			(33.12, 31.73)
			(33.06, 31.75)
			(33.06, 31.78)
			(33.06, 31.8)
			(33.06, 31.84)
			(33.06, 31.84)
		};
		\addlegendentry{$q=2$}

		\addplot[only marks, mark=oplus,black] plot coordinates {
	(33.14, 25.43) 
};
\addlegendentry{$tuning, q=2^{8}$}

		\addplot[only marks, mark=square,black] plot coordinates {
			(33.94, 25.13) 
		};
		\addlegendentry{$tuning, q=2^{4}$}
				\addplot[only marks, mark=diamond,black] plot coordinates {
			( 34.51, 24.82) 
		};
		\addlegendentry{$tuning, q=2$}
				\node[coordinate, pin=below:{w=1}]
		at (axis cs: 137.22, 49.42) {};
		
		\node[coordinate, pin=below:{w=2}]
		at (axis cs: 87.66, 38.78) {};
		
		\node[coordinate, pin=right:{w=4}]
		at (axis cs: 37.94, 29.07) {};
		
		\node[coordinate, pin=above:{RLNC}]
		at (axis cs: 33.06, 31.78) {};
		\end{axis}
		\end{tikzpicture}}
	\qquad
	\subfloat[$n=64$]{\begin{tikzpicture}
		[scale= 0.5]
		\begin{axis}[
		xlabel= $Number\; of\; Transmissions$,
		ylabel=$ADD$,
		xlabel style={font=\large},
		grid=major,
		legend pos= outer north east]
		
			\addplot[dotted, smooth,,black] plot coordinates {
			
	(266.09, 75.63) 
(154.23, 63.74) 
(88.94, 61.06)
(65.62, 62.84)
(64.58, 63.81)
(64.08, 63.87)
(64.06, 63.89)
(64.06, 63.91)
(64.06, 63.91)
(64.06, 63.91)
(64.06, 63.91)
(64.06, 63.91)
(64.06, 63.91)
(64.06, 63.91)
(64.06, 63.91)
(64.06, 63.91)
			
		};
		\addlegendentry{$q=2^{8}$}
\addplot[smooth,mark=-, red] plot coordinates {
		(267.18, 74.59) 
		(157.26, 63.67) 
		(89.73, 60.86)
		(65.82, 62.63)
		(64.73, 63.74)
		(64.09, 63.82)
		(64.08, 63.84)
		(64.08, 63.89)
		(64.08, 63.89)
		(64.08, 63.89)
		(64.08, 63.89)
		(64.08, 63.89)
		(64.08, 63.89)
		(64.08, 63.89)
		(64.08, 63.89)
		(64.08, 63.89)
	};
	\addlegendentry{$q=2^{4}$}
		\addplot[dashed, smooth,mark=+,blue] plot coordinates {
			
			(269.56, 73.76) 
			(159.4, 63.53) 
			(91.34, 60.27)
			(67.31, 61.64)
			(66.22, 62.73)
			(65.66, 62.94)
			(65.63, 63.57)
			(65.61, 63.79)
			(65.6, 63.82)
			(65.6, 63.82)
			(65.6, 63.82)
			(65.6, 63.82)
			(65.6, 63.82)
			(65.6, 63.82)
			(65.59, 63.82)
			(65.59, 63.82)
					
		};
		\addlegendentry{$q=2$}

	\addplot[only marks, mark=oplus,black] plot coordinates {
	( 64.32, 51.36) 
};
\addlegendentry{$tuning, q=2^{8}$}
		\addplot[only marks, mark=square,black] plot coordinates {
	( 64.58, 50.57) 
};
\addlegendentry{$tuning, q=2^{4}$}
				\addplot[only marks, mark=diamond,black] plot coordinates {
			( 66.18, 49.57) 
		};
		\addlegendentry{$tuning, q=2$}

				\node[coordinate, pin=below:{w=1}]
		at (axis cs: 269.56, 73.76) {};
		
		\node[coordinate, pin=below:{w=2}]
		at (axis cs: 159.4, 63.53) {};
		
		\node[coordinate, pin=below:{w=4}]
		at (axis cs: 91.34, 60.27) {};
		
		\node[coordinate, pin=above:{RLNC}]
		at (axis cs: 65.59, 63.82) {};
				\end{axis}
		\end{tikzpicture}}
	\caption[2]{Average ADD and the number of required transmissions for Tuning scheme and traditional SNC for $n=32,64$ and $q=2,2^{4},2^{8}$}
	\label{fig:4}
\end{figure*}

\begin{figure*}[!t]
	\centering
	\subfloat[$n =32$]{\begin{tikzpicture}
		[scale= 0.55]
		\begin{axis}[
		xlabel= $w$,
		ylabel=$Number\; of\; Transmissions$,
		xlabel style={font=\large},
		grid=major,
		legend pos= outer north east]
		

		\addplot[smooth,mark=-,dashed, red] plot coordinates {
			(3, 46.22)
			(4, 36.59)
			(5, 33.82)
			(6, 33.77)
			(7, 33.59)
			(8, 33.24)
			(9, 33.08)
			(10, 32.63)
			(11, 32.63)
			(12, 32.63)
			(13, 32.63)
			(14, 32.63)
			(15, 32.63)
			(16, 32.63)
		};
		\addlegendentry{$q=2^{4}$}
		
		\addplot[smooth,mark=+,blue] plot coordinates {
			(3, 47.44)
			(4, 37.94)
			(5, 34.79)
			(6, 34.12)
			(7, 33.86)
			(8, 33.73)
			(9, 33.48)
			(10, 33.26)
			(11, 33.12)
			(12, 33.06)
			(13, 33.06)
			(14, 33.06)
			(15, 33.06)
			(16, 33.06)
		};
		\addlegendentry{$q=2$}
		

%
		
		\addplot[mark=square,black] plot coordinates {
		(3, 33.94)
		(4, 33.94)
		(5, 33.94)
		(6, 33.94)
		(7, 33.94)
		(8, 33.94)
		(9, 33.94)
		(10, 33.94)
		(11, 33.94)
		(12, 33.94)
		(13, 33.94)
		(14, 33.94)
		(15, 33.94)
		(16, 33.94)
		};
		\addlegendentry{$tuning, q=2^{4}$}
		\addplot[mark=diamond,black] plot coordinates {
		(3, 34.51)
		(4, 34.51)
		(5, 34.51)
		(6, 34.51)
		(7, 34.51)
		(8, 34.51)
		(9, 34.51)
		(10, 34.51)
		(11, 34.51)
		(12, 34.51)
		(13, 34.51)
		(14, 34.51)
		(15, 34.51)
		(16, 34.51)

		};
		\addlegendentry{$tuning, q=2$}
		\end{axis}
		\end{tikzpicture}}
	\qquad
	\subfloat[$n=64$]{\begin{tikzpicture}
		[scale= 0.55]
		\begin{axis}[
		xlabel= $w$,
		ylabel=$Number\; of\; Transmissions$,
		xlabel style={font=\large},
		grid=major,
		legend pos= outer north east]
		
		%
		%

		\addplot[smooth,mark=-, red] plot coordinates {
			(3, 89.73)
			(4, 65.82)
			(5, 64.73)
			(7, 64.09)
			(9, 64.08)
			(11, 64.08)
			(13, 64.08)
			(15, 64.08)
			(17, 64.08)
			(20, 64.08)
			(23, 64.08)
			(26, 64.08)
			(29, 64.08)
			(31, 64.08)
		};
		\addlegendentry{$q=2^{4}$}

		\addplot[dashed, smooth,mark=+,blue] plot coordinates {
			
			(3, 91.34)
			(4, 67.31)
			(5, 66.22)
			(7, 65.66)
			(9, 65.63)
			(11, 65.61)
			(13, 65.6)
			(15, 65.6)
			(17, 65.6)
			(20, 65.6)
			(23, 65.6)
			(26, 65.6)
			(29, 65.59)
			(31, 65.59)
		};
		\addlegendentry{$q=2$}
		
		\addplot[ mark=square,black] plot coordinates {
		(3, 64.58)
		(4, 64.58)
		(5, 64.58)
		(7, 64.58)
		(9, 64.58)
		(11, 64.58)
		(13, 64.58)
		(15, 64.58)
		(17, 64.58)
		(20, 64.58)
		(23, 64.58)
		(26, 64.58)
		(29, 64.58)
		(31, 64.58)
		};
		\addlegendentry{$tuning, q=2^{4}$}
		\addplot[mark=diamond,black] plot coordinates {
		(3, 66.18)
		(4, 66.18)
		(5, 66.18)
		(7, 66.18)
		(9, 66.18)
		(11, 66.18)
		(13, 66.18)
		(15, 66.18)
		(17, 66.18)
		(20, 66.18)
		(23, 66.18)
		(26, 66.18)
		(29, 66.18)
		(31, 66.18)
		};
		\addlegendentry{$tuning, q=2$}
		
		\end{axis}
		\end{tikzpicture}}
	\caption[2]{Average number of transmissions for tuning scheme and traditional SNC for $n=32,64$ and $q=2,2^{4}$}
	\label{fig:6}
\end{figure*}
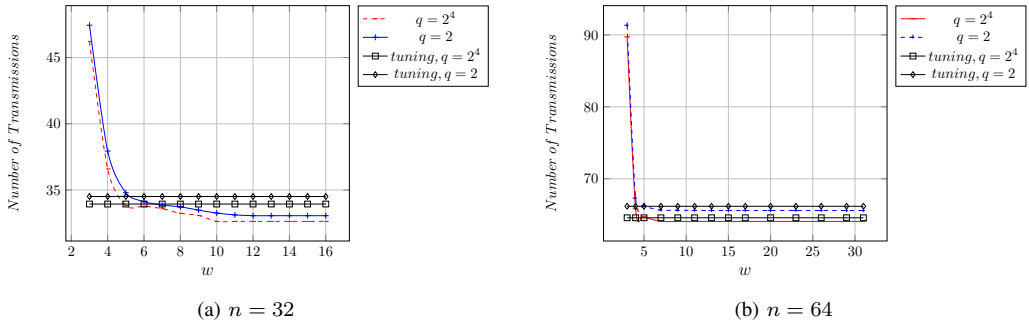

The results of our tuning scheme are compared to traditional SNC in fig.~\ref{fig:4} to Fig.~\ref{fig:6}. These figures show the impact of the tuning scheme on ADD and the average number of transmissions where $n\in \{32,64\}$ and $q\in \{2,2^{4},2^{8}\}$. As these figures suggest, our tuning scheme achieves an average 21~\% lower ADD while having an average of 1~\% more transmissions with respect to traditional RLNC. The ADD of the proposed tuning scheme is also 16~\% better than the best achievable ADD with traditional SNC, regardless of the number of source packets and Galois Field. Table~\ref{tab:1} also summarizes the results for the average ADD and the average number of transmissions for different values of sparsity and our tuning scheme. In this table we show the values for the ADD and the Average Number of Transmissions (ANT) for different values of sparsity in traditional SNC and RLNC as well as our tuning scheme for different values of $n$ and $q$. In this table we also show the difference of ADD and ANT in tuning scheme and traditional RLNC to better illustrate our achievement.

\begin{figure*}[!t]
	\centering
	\subfloat[$n =32$]{\begin{tikzpicture}
		[scale= 0.55]
		\begin{axis}[
		xlabel= $w$,
		ylabel=$ADD$,
		xlabel style={font=\large},
		grid=major,
		legend pos= outer north east]
		
		
		\addplot[smooth,mark=-,dashed, red] plot coordinates {
			(1, 49.76) 
			(2, 38.78) 
			(3, 30.46)
			(4, 29.18)
			(5, 30.83)
			(6, 31.66)
			(7, 31.72)
			(8, 31.76)
			(9, 31.77)
			(10, 31.78)
			(11, 31.81)
			(12, 31.83)
			(13, 31.84)
			(14, 31.84)
			(15, 31.84)
			(16, 31.84)
		};
		\addlegendentry{$q=2^{4}$}
		
		\addplot[smooth,mark=+,blue] plot coordinates {
			(1, 49.42) 
			(2, 38.78) 
			(3, 30.37)
			(4, 29.07)
			(5, 30.48)
			(6, 31.48)
			(7, 31.56)
			(8, 31.66)
			(9, 31.69)
			(10, 31.71)
			(11, 31.73)
			(12, 31.75)
			(13, 31.78)
			(14, 31.8)
			(15, 31.84)
			(16, 31.84)
		};
		\addlegendentry{$q=2$}
		
		
		\addplot[mark=square,black] plot coordinates {
		(1, 25.13) 
		(2, 25.13) 
		(3, 25.13)
		(4, 25.13)
		(5, 25.13)
		(6, 25.13)
		(7, 25.13)
		(8, 25.13)
		(9, 25.13)
		(10, 25.13)
		(11, 25.13)
		(12, 25.13)
		(13, 25.13)
		(14, 25.13)
		(15, 25.13)
		(16, 25.13)
		};
		\addlegendentry{$tuning, q=2^{4}$}
		\addplot[mark=diamond,black] plot coordinates {
			(1, 24.82) 
			(2, 24.82) 
			(3, 24.82)
			(4, 24.82)
			(5, 24.82)
			(6, 24.82)
			(7, 24.82)
			(8, 24.82)
			(9, 24.82)
			(10, 24.82)
			(11, 24.82)
			(12, 24.82)
			(13, 24.82)
			(14, 24.82)
			(15, 24.82)
			(16, 24.82)
		};
		\addlegendentry{$tuning, q=2$}
		\end{axis}
		\end{tikzpicture}}
	\qquad
	\subfloat[$n=64$]{\begin{tikzpicture}
		[scale= 0.55]
		\begin{axis}[
		xlabel= $w$,
		ylabel=$ADD$,
		xlabel style={font=\large},
		grid=major,
		legend pos= outer north east]
		
%
%
		\addplot[smooth,mark=-, dashed, red] plot coordinates {
			(1, 74.59) 
			(2, 63.67) 
			(3, 60.86)
			(4, 62.63)
			(5, 63.74)
			(7, 63.82)
			(9, 63.84)
			(11, 63.89)
			(13, 63.89)
			(15, 63.89)
			(17, 63.89)
			(20, 63.89)
			(23, 63.89)
			(26, 63.89)
			(29, 63.89)
			(31, 63.89)
		};
		\addlegendentry{$q=2^{4}$}
		\addplot[ smooth,mark=+,blue] plot coordinates {
			
			(1, 73.76) 
			(2, 63.53) 
			(3, 60.27)
			(4, 61.64)
			(5, 62.73)
			(7, 63.1  4)
			(9, 63.57)
			(11, 63.79)
			(13, 63.82)
			(15, 63.82)
			(17, 63.82)
			(20, 63.82)
			(23, 63.82)
			(26, 63.82)
			(29, 63.82)
			(31, 63.82)
		};
		\addlegendentry{$q=2$}
		
		\addplot[ mark=square,black] plot coordinates {
					(1, 50.57) 
		(2, 50.57) 
		(3, 50.57)
		(4, 50.57)
		(5, 50.57)
		(7, 50.57)
		(9, 50.57)
		(11, 50.57)
		(13, 50.57)
		(15, 50.57)
		(17, 50.57)
		(20, 50.57)
		(23, 50.57)
		(26, 50.57)
		(29, 50.57)
		(31, 50.57)
		};
		\addlegendentry{$tuning, q=2^{4}$}
		\addplot[mark=diamond,black] plot coordinates {
						(1, 49.57) 
			(2, 49.57) 
			(3, 49.57)
			(4, 49.57)
			(5, 49.57)
			(7, 49.57)
			(9, 49.57)
			(11, 49.57)
			(13, 49.57)
			(15, 49.57)
			(17, 49.57)
			(20, 49.57)
			(23, 49.57)
			(26, 49.57)
			(29, 49.57)
			(31, 49.57)
		};
		\addlegendentry{$tuning, q=2$}
		
		\end{axis}
		\end{tikzpicture}}
	\caption[2]{Average ADD for Tuning scheme and traditional SNC for $n=32,64$ and $q=2,2^{4}$}
	\label{fig:5}
\end{figure*}
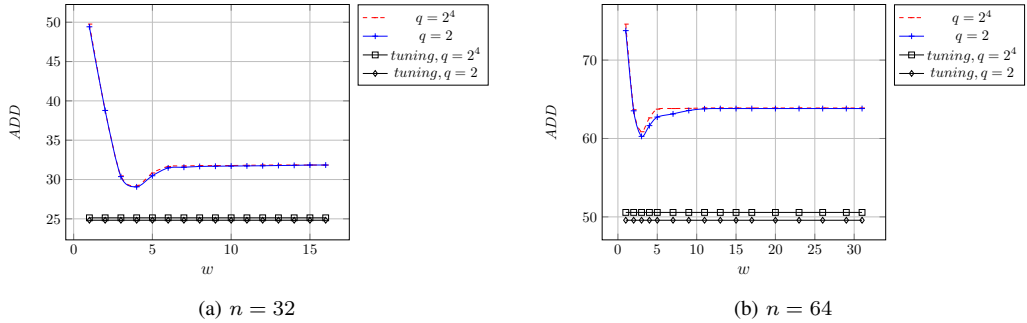

%
\section{Conclusion}
\label{conclusion}

In this paper, we derived an approximation for the probability of partial decoding in SNC tuning scheme. Our results show the proposed model predicts the fraction of decoded source packets after an arbitrary number of transmissions with an average RMSD of 6~\% regardless of sparsity and Galois Field. Our results show that at the start of transmission, higher sparsity, increase the probability of partial decoding. However, towards the end of transmission, the probability of partial decoding decreases for higher sparsity and increases for lower sparsity. Therefore, if the sparsity of system is constant throughout the whole transmission, the partial decoding will perform as desired in only a part of the whole transmission.

\begin{table*}[!t]
\renewcommand{\arraystretch}{0.8}
\scriptsize
\caption{Average ADD and average ANT for tuning scheme, traditional SNC and RLNC for different values of $n$ and $q$}
\label{tab:1}
\centering
\tiny{
\begin{tabular}{|c|c|c||c|c|c|c|c|c||c|c|}
\hline
Generation & Galois Field & & $w=1$ & $w=2$ & $w=3$ &$w=5$ & $w=10$ & RLNC & Tuning & Improvement\\
\hline
\multirow{2}{*}{32} & \multirow{2}{*}{2} & Average Decoding Delay & 48.76 & 37.78 & 30.37 & 30.48 & 31.71 & 31.84 & 25.43 & -20.13\% \\
 & & Average \# Transmissions & 137.22 & 87.66 & 47.44 & 34.79 & 33.26 & 33.06 & 33.94 & + 2.66\%\\
 \hline
 \multirow{2}{*}{32} & \multirow{2}{*}{$2^{4}$} & Average Decoding Delay & 49.42 & 38.78 & 30.46 & 30.83 & 31.8 &  31.84 & 25.13 & -21.07\% \\
 & & Average \#  Transmissions & 135.91 & 84.73 & 46.22 & 33.82 & 32.63 & 32.63 & 32.81 & +0.55\%\\
 \hline
  \multirow{2}{*}{32} & \multirow{2}{*}{$2^{8}$} & Average Decoding Delay & 50.16 & 39.64 & 30.73 & 31.03 & 31.81 & 31.84 & 24.83& -22.01\% \\
 & & Average \#  Transmissions & 133.76 & 82.31 & 45.84 & 33.59 & 31.58 & 32.58 & 32.61 & +0.1\% \\
 \hline
  \multirow{2}{*}{64} & \multirow{2}{*}{2} & Average Decoding Delay & 73.76 & 63.53 & 59.86 & 62.73 & 63.68 & 63.68 & 49.57 & -22.15\% \\
 & & Average \# Transmissions & 269.56 & 159.4 & 91.34 & 66.22 & 65.62 & 65.62 & 66.18 & +0.85\% \\
 \hline
   \multirow{2}{*}{64} & \multirow{2}{*}{$2^{4}$} & Average Decoding Delay & 74.59 & 63.67 & 60.86 & 63.74 & 63.87 & 63.87 & 50.57 &  -21.29\% \\
 & & Average \# Transmissions & 267.18 & 157.26 & 89.73 & 64.73 & 64.08 & 64.08 & 64.58 &+0.77\% \\
 \hline
    \multirow{2}{*}{64} & \multirow{2}{*}{$2^{8}$} & Average Decoding Delay & 75.63 & 63.74 & 61.06 & 63.81 & 63.91 & 63.91 & 51.36& -19.63\% \\
 & & Average \# Transmissions & 266.09 & 154.23 & 88.94 & 64.58 & 64.06 & 64.06 & 64.32 & +0.37\% \\
 \hline
\end{tabular}}
\end{table*}

In order to achieve the maximum partial decoding probability throughout the whole transmission, we proposed a tuning scheme. In this tuning scheme, we change the sparsity in each coded packet. The results show this tuning scheme achieves an average 21~\% lower ADD compared to RLNC and an average 16~\% lower ADD with respect to best possible ADD for SNC scheme with fixed sparsity. The proposed tuning scheme also manages to maintain the average number of required transmissions close to traditional RLNC, with an average of 1.5~\% more required transmissions.

\appendix
\section{Proof of Theorem 1.}
 Consider the source packets as $(e_{1}, e_{2}, ..., e_{n})$. In this notation, a coded packet can be considered as a linear combination of $w$ source packets, where $w = p\times n$. The probability that at least $x$ source packets are decoded in the decoder is equal to the probability that for a set $S\subseteq \{1,2,...,n\}, |S| = x$, we have $\forall i \in S: e_{i}\in Row(M)$. let's denote this probability as $g(S)$. Also, we denote $f(S)$ as the probability that exactly the source packets enumerated in $S$ are decoded. This probability is equal to the probability that  $\forall i \in S: e_{i}\in Row(M)$ and  $\forall j \notin S: e_{j}\notin Row(M)$.  It is easy to see that $g(S) = \sum_{J: J\supseteq S}f(J)$. By using lemmas~1 and~2, for any set $S$, we have:
\begin{equation}
g(S) = \frac{N_{|S|}.P_{m-|S|}^{r-|S|}}{NP_{m}^{r}}.
\end{equation}

The terms $P_{m-|S|}^{r-|S|}$ and $P_{m-|S|}^{r-|S|}$ are added to enumerate the number of matrices with rank $r-|S|$ and $r$ respectively. Provoking the principal of Inclusion and Exclusion, we get the results of ~\eqref{eq:res1} for $f(S)$.

\begin{figure*}[!t]
	\normalsize
	\setcounter{equation}{30}
\begin{equation}
\begin{split}
&f(S) = \sum_{J: J\supseteq S}(-1)^{|J/S|} \frac{N_{|J|}P_{m-|J|}^{r-|J|}}{nP_{m}^{r}} = \\
&\sum_{\substack{J: J\supseteq S \\ |J| \leq m}}(-1)^{|J/S|}\frac{\big(\frac{\Gamma(n+1)}{\Gamma(np_{|J|}+1)\Gamma(n-np_{|J|}+1)}(q-1)^{np_{|J|}}\big)^{m-|J|}P_{m-|J|}^{r-|J|}}{nP_{m}^{r}} \\
&= \sum_{\substack{J^{\prime}: J^{\prime}\subseteq \{1,...,n\}/S \\ |J^\prime| \leq m-x}}(-1)^{|J^{\prime}|}\frac{\big(\frac{\Gamma(n+1)}{\Gamma(np_{x +|J^{\prime}|}+1)\Gamma(n-np_{x+|J^{\prime}|}+1)}(q-1)^{np_{x+|J^{\prime}|}}\big)^{m-x-|J^{\prime}|}P_{m-x-|J^{\prime}|}^{r-x-|J^{\prime}|}}{nP_{m}^{r}}.
\end{split}
\label{eq:res1}
\end{equation}
	\setcounter{equation}{31}
	\hrulefill
	\vspace*{4pt}
\end{figure*}

The last equality in~\eqref{eq:res1} follows by setting the value of $J^{\prime} = J/S$. 

There are ${n-x\choose j}$ possible sets of $J^{\prime}$, where $|J^{\prime}| = j$, hence,~\eqref{eq:res1} takes following form.
\begin{equation}
\begin{split}
&f(S) =  \sum_{j=0}^{m-x}{n-x\choose j} (-1)^{j} \\ &\frac{\big(\frac{\Gamma(n+1)}{\Gamma(np_{x +j}+1)\Gamma(n-np_{x+j}+1)}(q-1)^{np_{x+j}}\big)^{m-x-j}P_{m-x-j}^{r-x-j}}{nP_{m}^{r}}.
\end{split}
\label{eq:res2}
\end{equation}

Considering that $f(S)$ is the probability that exactly the source packets indicated by $S$ are decoded, we can write

\begin{equation}
P(X = x|R=r,M=m) = \sum_{S:|S|=x}f(S) = {n\choose x}f(S^{\prime})
\label{eq:res3}
\end{equation}

where $S^{\prime}$ is any subset of $\{1,2,...,k\}$ of size $x$. Substituting~\eqref{eq:res2} in~\eqref{eq:res3} gives the result. 

\ifCLASSOPTIONcaptionsoff
  \newpage
\fi

\end{document}